\documentclass[twocolumn]{aastex61}

\received{2017 December 22}
\revised{2018 March 09}
\accepted{2018 March 31}
\usepackage{graphicx}
\usepackage{amssymb}
\usepackage{amsmath}
\usepackage{epstopdf}
\shorttitle{An Empirical Planetesimal Belt Radius - Stellar Luminosity Relation}
\shortauthors{Matr\`a et al.}

\begin{document}


\title{An Empirical Planetesimal Belt Radius - Stellar Luminosity Relation}

\author[0000-0003-4705-3188]{L. Matr\`a}
%\altaffiltext{1}{Institute of Astronomy, University of Cambridge, Madingley Road, Cambridge CB3 0HA, UK}
\altaffiliation{Submillimeter Array (SMA) Fellow}
\affil{Harvard-Smithsonian Center for Astrophysics, 60 Garden Street, Cambridge, MA 02138, USA}
%\affil{Institute of Astronomy, University of Cambridge, Madingley Road, Cambridge CB3 0HA, UK}
\email{luca.matra@cfa.harvard.edu}
\author{S. Marino}
\affil{Institute of Astronomy, University of Cambridge, Madingley Road, Cambridge CB3 0HA, UK}
%\author{friends}
\author{G. M. Kennedy}
\affil{Department of Physics, University of Warwick, Gibbet Hill Road, Coventry, CV4 7AL, UK}
\author{M. C. Wyatt}
\affil{Institute of Astronomy, University of Cambridge, Madingley Road, Cambridge CB3 0HA, UK}
\author{K. I. \"{O}berg}
\affil{Harvard-Smithsonian Center for Astrophysics, 60 Garden Street, Cambridge, MA 02138, USA}
\author{D. J. Wilner}
\affil{Harvard-Smithsonian Center for Astrophysics, 60 Garden Street, Cambridge, MA 02138, USA}

\begin{abstract}
Resolved observations of millimetre-sized dust, tracing larger planetesimals, have pinpointed the location of 26 Edgeworth-Kuiper belt analogs. 
We report that a belt's distance $R$ to its host star correlates with the star's luminosity $L_{\star}$, following $R\propto L^{0.19}_{\star}$ with a low intrinsic scatter of $\sim$17\%. Remarkably, our Edgeworth-Kuiper belt in the Solar System and the two CO snow lines imaged in protoplanetary disks lie close to this $R$-$L_{\star}$ relation, suggestive of an intrinsic relationship between protoplanetary disk structures and belt locations. %Other explanations for this coincidence exist as well, however.
To test the effect of bias on the relation, we use a Monte Carlo approach and simulate uncorrelated model populations of belts. We find that observational bias could produce the slope and intercept of the $R$-$L_{\star}$ relation, but is unable to reproduce its low scatter.
We then repeat the simulation taking into account the collisional evolution of belts, following the steady state model that fits the belt population as observed through infrared excesses. This significantly improves the fit by lowering the scatter of the simulated $R$-$L_{\star}$ relation; however, this scatter remains only marginally consistent with the one observed.
The inability of observational bias and collisional evolution alone to reproduce the tight relationship between belt radius and stellar luminosity could indicate that planetesimal belts form at preferential locations within protoplanetary disks. The similar trend for CO snow line locations would then indicate that the formation of planetesimals and/or planets in the outer regions of planetary systems is linked to the volatility of their building blocks, as postulated by planet formation models.
%We conclude by discussing this and other implications that this relation may have on the timescale and/or the location of planetesimal and planet formation in the outer regions of planetary systems. 

\end{abstract}


%% subject header.  Objects should be in the appropriate "individual"
%% headers (e.g. quasars: individual, stars: individual, etc.) with the
%% additional provision that the total number of headers, including each
%% individual object, not exceed six.  The \objectname{} macro, and its
%% alias \object{}, is used to mark each object.  The macro takes the object
%% name as its primary argument.  This name will appear in the paper
%% and serve as the link's anchor in the electronic edition if the name
%% is recognized by the data centers.  The macro also takes an optional
%% argument in parentheses in cases where the data center identification
%% differs from what is to be printed in the paper.

%\keywords{globular clusters: general ---
%globular clusters: individual(\objectname{NGC 6397},
%\object{NGC 6624}, \objectname[M 15]{NGC 7078},
%\object[Cl 1938-341]{Terzan 8})}



\section{Introduction}
\label{sect:intro}

The ubiquity of gas-poor, optically thin dust disks, known as \textit{debris disks}, around main sequence stars tells us that belts of planetesimals are a likely outcome of the planet formation process \citep[for a review, see][]{Matthews2014}. Planetesimal belts may form in the younger, dust- and gas-rich environments of protoplanetary disks, where the bulk of planet formation is thought to take place, but may also be produced after gas dispersal as a by-product of terrestrial planet formation. 
Formation in protoplanetary disks is likely for extrasolar Kuiper belts in the outer regions of planetary systems, as indicated by the increasing number of detections of large amounts of gas %- at times comparable to protoplanetary disks - 
in young ($\lesssim$ few tens of Myr), cold ($\gtrsim$ 10 au) debris disks \citep[e.g.][]{Greaves2016,Lieman-Sifry2016,Moor2017}.
However, why and how planetesimal belts arise remains largely unknown, and observations of individual systems provide few constraints on this process \citep{Wyatt2015}. 

One aspect of planetesimal belts that can be linked to their formation mechanism is their location, %As opposed to their mass, which depletes over time due to collisional evolution \citep[e.g.][]{Wyatt2002, Rieke2005}, their location 
which should remain unchanged over long timescales once the planets have formed and settled to a stable configuration. This is particularly true given that the observed evolution of belt masses (at least around A stars)
argues for the majority of the belt population being narrow rings \citep{Kennedy2010}. 
The presence of a planetesimal belt in a planetary system tells us that, at that location, grain growth must have been efficient enough to form planetesimals, although some mechanism must have also been in place to either prevent further growth into planets, or to remove planets from these regions fast enough to produce a second generation of planetesimals before the gas-rich protoplanetary disk dissipated. Can these conditions arise anywhere in a planetary system? Or are there specific locations where these mechanisms giving rise to planetesimal belts preferentially act? 
%Thus, the location of planetesimal belts as a population could yield important insights on their formation mechanism.

%nother important aspect that we can address through studying the location of planetesimal belts is their composition. The composition of forming planetesimals is a function of where these planetesimals formed, meaning that knowledge of a belt's formation location combined with chemical imaging of line emission in young protoplanetary disks may be able to indicate their most likely composition. Even better, an exciting prospect that has recently emerged is that of measuring exocometary volatile compositions \citep{Matra2015,Matra2017a,Marino2016,Matra2017c}.
%, which can then be linked to the belt's location and line observations of young disks to study the origin of cometary ices. 
%For example, the exocometary CO(+CO$_2$) content of exocomets in nearby planetesimal belts appears to be similar to that of Solar System comets \citep{Matra2017b}, potentially indicating that belts in different planetary systems formed in similar conditions - perhaps at similar locations - within protoplanetary disks. 

%Thus, studying the location of planetesimal belts as a population will yield important insights on their formation mechanism. 
Current planet and planetesimal formation theories predict that planet and/or planetesimal formation efficiency is a function of distance to the central star. In the core accretion scenario, this naturally arises from timescale and temperature arguments \citep[e.g.][]{Lissauer1987, Lewis1974}. We focus on distances of tens of au, where most known planetesimal belts are observed. Growth timescales increase further away from the star, so for a given protoplanetary disk lifetime, planets may only have enough time to form out to a certain distance, leaving a planetesimal belt beyond. At the same time, temperatures decrease with distance to the central star, creating several compositional transitions, or \textit{snow lines}, beyond which gas species of increasing volatility can freeze out onto solid grains \citep[e.g.][]{Cuzzi2004}. This can affect growth in different ways, for example through the sticking and fragmentation efficiency of particles \citep[e.g.][]{Wada2009, Okuzumi2016}, but also by creating pressure gradients in the gas affecting particle concentrations \citep[e.g.][]{Stevenson1988}.
In general, theory would therefore suggest that the presence of a planetesimal belt, be it caused by failed growth to planets or enhanced planetesimal formation, should relate to distance to the central star. This motivates studies that observationally constrain the location of planetesimal belts as a population, and that test its dependence with host star properties - such as mass and luminosity - which directly affect the radial dependence of planet and/or planetesimal formation efficiency.

%In order to answer these questions and gain insights on their formation mechanism, studying the location of planetesimal belts as a population is necessary. 

Such studies have so far been limited by the fact that, for the vast majority of belts, we only have unresolved IR multiband photometry constraining the dust temperature $T$ of the small grains. This gives us a rough idea of a belt's location under the assumption that the grains emit as blackbodies, giving us their blackbody radius $R_{\rm BB}$. Several studies have analyzed the dependence of dust temperature on host star properties \citep[e.g.][]{Chen2014, Jang-Condell2015}, with \citet{Ballering2013} for example finding that the temperature of outer belts correlates with the temperature of the host star. However, it is well established that the small grains traced by the temperature of the spectral energy distribution are generally hotter than blackbody by an amount that is dependent on the grain properties and the size distribution; this means that $R_{\rm BB}$ only truly gives us a lower limit to a belt's location $R$ \citep[e.g.][]{Booth2013}. 

Studies such as that of \citet{Ballering2017} alleviate this effect by accounting for the dust's optical properties, assuming all belts share the same composition, and finding that the radial location of warm, inner belts increases around stars with increasing masses, once again with a large scatter. %This goes in the same direction as previous results finding that the temperature of cold, outer belts positively correlates with the temperature of the host star \citep{Ballering2013}, with the trend being steeper than linear \citep{KennedyWyatt2014}. 
%Motivated to understand their formation mechanism and its compositional implications, the goal of this work is to use resolved imaging to unambiguously determine the location of planetesimal belts as a population. 
In addition, \textit{Herschel} marginally resolved a considerable number of cold dust disks \citep[e.g.][]{Morales2016, Kennedy2015, Moor2015} mostly at 70-100$\mu$m wavelengths where its resolution was 5$\arcsec$-7$\arcsec$, corresponding to 100-700 au at distances between 20 and 100 pc from Earth where the bulk of the observed population lies. However, \textit{Herschel} studies were limited by 1) the accuracy of radius determination, due to the poor spatial resolution and stellar emission contaminating the disk's inner regions, 2) observational bias due to the inability to resolve disks smaller than the resolution quoted above, and 3) the fact that IR observations probe small grains that are dynamically affected by radiation forces \citep[e.g.][]{Burns1979, StrubbeChiang2006} and may therefore not directly trace the location of the parent planetesimals.

A solution to these issues is to resolve belts through millimetre wavelength interferometry, where the star is in most cases too faint to be detected, the resolution is sufficiently high to resolve even the smallest disks, and mm-sized grains are not subject to radiation forces, remaining in the same low eccentricity orbits as the planetesimals they are created from \citep[e.g.][]{Wyatt2006}. We here present a first population study of planetesimal belt locations derived through millimeter wavelength interferometric observations.
In \S\ref{sect:res} we introduce the full sample of interferometrically resolved planetesimal belts from the literature, showing that the distance of a belt from its host star (i.e. its radius) correlates with the star's luminosity. In \S\ref{sect:anal} we qualitatively and quantitatively analyse the impact of observational bias, showing its importance in assessing the nature of correlations obtained from biased datasets. Having established the likely presence of an underlying correlation, in \S\ref{sect:disc} we proceed to interpret the correlation in the context of both the collisional evolution and the formation location of planetesimal belts, and consider its potential implications for planetesimal and planet formation at large orbital separations. We conclude with a summary of our findings in \S\ref{sec:concl}.

\section{Results}
\label{sect:res}

\begin{table*}[h]
\begin{center}
\caption{Properties of the sample of planetesimal belts resolved at mm wavelengths}
\label{tab:sourcelist}
\begin{tabular}{ c c c c c c c c c c c c}
\hline
\hline % (L$_{\odot}$) 
HD name & Name & $d$ & SpT & $L_{\star}$ & $M_{\star}$ & Age & $R$ & $\Delta R$ & $R_{\rm BB}$ & $f$ & Ref. \\ 
& & (pc) & & ($L_{\odot}$) & ($M_{\odot}$) & (Myr) & (au) & (au) & (au) & & \\
\hline
HD 377 &  & 39.1 & G2V & 1.2 & 1.1 & 220 & 63.0 & 32.0 & 31.4 & 3.6e-04 & 1 \\
HD 8907 &  & 34.8 & F8 & 2.0 & 1.2 & 200 & 80.0 & 52.0 & 46.5 & 2.3e-04 & 1 \\
HD 9672 & 49 Ceti & 59.4 & A1V & 15.8 & 1.9 & 40 & 228.0 & 310.0 & 85.4 & 7.2e-04 & 2 \\
HD 10700 & $\tau$ Ceti & 3.7 & G8.5V & 0.5 & 0.9 & 5800 & 29.1 & 45.8 & 7.0 & 1.3e-05 & 3 \\
HD 15115 &  & 45.2 & F4IV & 3.6 & 1.3 & 23 & 78.2 & 69.6 & 55.1 & 4.6e-04 & 4 \\
HD 21997 &  & 71.9 & A3IV/V & 9.9 & 1.7 & 30 & 106.0 & 88.0 & 65.4 & 5.6e-04 & 5 \\
HD 22049 & $\epsilon$ Eri & 3.2 & K2Vk: & 0.3 & 0.8 & 600 & 69.4 & 11.4 & 19.5 & 4.0e-05 & 6 \\
HD 39060 & $\beta$ Pic & 19.4 & A6V & 8.1 & 1.6 & 23 & 100.0 & 100.0 & 24.3 & 2.1e-03 & 7 \\
HD 61005 &  & 35.3 & G8Vk & 0.7 & 0.9 & 40 & 66.4 & 23.6 & 21.0 & 2.3e-03 & 8 \\
HD 95086 &  & 90.4 & A8III & 6.1 & 1.7 & 15 & 204.0 & 176.0 & 46.5 & 1.4e-03 & 17 \\
HD 104860 &  & 45.5 & F8 & 1.2 & 1.0 & 250 & 164.0 & 108.0 & 44.5 & 5.3e-04 & 1 \\
HD 107146 &  & 27.5 & G2V & 1.0 & 1.0 & 200 & 88.6 & 126.8 & 37.8 & 8.6e-04 & 9 \\
HD 109085 & $\eta$ Crv & 18.3 & F2V & 5.0 & 1.4 & 1400 & 152.0 & 46.0 & 52.9 & 2.9e-05 & 10 \\
HD 111520 &  & 108.6 & F5/6V & 3.0 & 1.3 & 15 & 96.0 & 90.0$^{\star}$ & 58.5 & 1.1e-03 & 11 \\
HD 115617 & 61 Vir & 8.6 & G7V & 0.8 & 1.0 & 6300 & 91.5 & 123.0 & 22.2 & 2.4e-05 & 12 \\
HD 121617 &  & 128.2 & A1V & 17.3 & 1.9 & 16 & 82.5 & 54.8 & 30.0 & 4.9e-03 & 18 \\
HD 131488 &  & 147.7 & A1V & 13.1 & 1.8 & 16 & 84.0 & 44.0 & 35.6 & 2.2e-03 & 18 \\
HD 131835 &  & 122.7 & A2IV & 11.4 & 2.0 & 16 & 91.0 & 140.0 & 57.0 & 2.2e-03 & 11 \\
HD 138813 &  & 150.8 & A0V & 16.7 & 2.2 & 10 & 105.0 & 70.0 & 69.6 & 6.0e-04 & 11 \\
HD 145560 &  & 133.7 & F5V & 3.2 & 1.4 & 16 & 88.0 & 70.0 & 22.0 & 2.1e-03 & 11 \\
HD 146181 &  & 146.2 & F6V & 2.6 & 1.3 & 16 & 93.0 & 50.0$^{\star}$ & 17.0 & 2.2e-03 & 11 \\
HD 146897 &  & 128.4 & F2/3V & 3.1 & 1.5 & 10 & 81.0 & 50.0$^{\star}$ & 15.6 & 8.2e-03 & 11 \\
HD 181327 &  & 51.8 & F6V & 2.9 & 1.3 & 23 & 86.0 & 23.2 & 50.1 & 2.1e-03 & 13 \\
HD 197481 & AU Mic & 9.9 & M1Ve & 0.1 & 0.6 & 23 & 24.6 & 31.6 & 11.9 & 3.3e-04 & 14 \\
HD 216956 & Fomalhaut & 7.7 & A4V & 16.1 & 1.9 & 440 & 143.1 & 13.6 & 72.2 & 7.5e-05 & 15 \\
HD 218396 & HR 8799 & 39.4 & F0V & 5.5 & 1.5 & 30 & 287.0 & 284.0 & 123.6 & 2.5e-04 & 16 \\

\hline
\end{tabular}
\ \\
\vspace{2mm}
Stellar luminosities $L_{\star}$, fractional luminosities $f$ and blackbody radii $R_{\rm BB}$ obtained from spectral energy distribution (SED) fitting as described in \citet{KennedyWyatt2014}, except for $\epsilon$ Eri, for which we use the fractional luminosity and blackbody radius of the cold belt from \citet{Greaves2014}. Stellar masses $M_{\star}$ are derived assuming stars have reached the main-sequence, using tabulated values from \citet{PecautMamajek2013}, for all stars older than 20 Myr except low-mass AU Mic, which is still pre-main sequence and for which we adopt the mass value from \citet{Boccaletti2015}. For stars younger than 20 Myr, we use values from \citet{Pecaut2012} except for HD95086 \citep[where we adopt the value from][]{Meshkat2013} and HD138813 \citep{Hernandez2005}. For HD121617 and HD131488 we found no literature value, which led us to adopt main-sequence values after verifying that the stars are close to reaching the main sequence \citep[using tracks from][]{Baraffe2015}. Ages are derived, where possible, from membership to Sco-Cen subregions \citep{PecautMamajek2016}, $\beta$ Pic \citep{Mamajek2014}, Columba and Argus moving groups \citep{Zuckerman2011}. For HD377, HD8907, HD104860, HD107146, ages are from \citet{Sierchio2014} and references therein, for $\tau$ Ceti the age is from \citet{MamajekHillenbrand2008}, for $\epsilon$ Eri we adopt an average value in the range reported by \citet{Janson2015}, for $\eta$ Corvi the age is from \citet{Casagrande2011}, for 61 Vir from \citet{Valenti2005}, and for Fomalhaut from \citet{Mamajek2012}. References for belt radius measurements: 1) \citet{Steele2016}: uniform surface density as a function of radius assumed. 2) \citet{Hughes2017}: single power law model, $\gamma=-1.29$. 3) \citet{Macgregor2016b}: single power law model, $\gamma=-0.3$. 4) \citet{Macgregor2015a}: single power law model, $\gamma=-0.5$.  5) \citet{Moor2013}: single power law model, $\gamma=-2.4$. 6) \citet{Booth2017}: Gaussian model. 7) \citet{Dent2014}: deprojected non-parametric dust distribution. 8) \citet{Olofsson2016}: double power law model, $\Delta R$ measured as full width at half maximum (FWHM). 9) \citet{Ricci2015a}, single power law model, $\gamma=0.74$. 10) \citet{Marino2017a}, Gaussian model. 11) \citet{Lieman-Sifry2016}, single power law model with $\gamma=-1.0$ assumed ($\Delta R$ values marked by $^{\star}$ were reported as upper limits). 13) \citet{Marino2017b}, single power law model, $\gamma=0.1$. 13) \citet{Marino2016}, Gaussian model. 14) \citet{Macgregor2013}, single power law model, $\gamma=2.3$. 15) \citet{Macgregor2017}, eccentric ring model. 16) \citet{Booth2016}, single power law model, $\gamma=-1.0$. 17) \citet{Su2017}, Gaussian model. 18) \citet{Moor2017}, Gaussian model.
\end{center}
\end{table*}

We collected all resolved Submillimeter Array (SMA) and Atacama Large Millimeter/submillimeter Array (ALMA) interferometric observations of planetesimal belts at millimetre/submillimetre wavelength published to date, to form a final sample of 26. Table \ref{tab:sourcelist} shows their belt and host star properties, as obtained from resolved observations in the literature and spectral energy distribution (SED) fitting \citep[where the latter constrained the stellar luminosity, blackbody radius, and belt fractional luminosity, as described in][]{KennedyWyatt2014}. We note that, for the less well resolved objects in our literature sample, SED and visibility fitting were used simultaneously to constrain the disk's surface density distribution \citep[e.g.][]{Steele2016}.

We choose the belt location (radius) $R$ to be represented by either the average between the best-fit inner and outer belt radii (for models with a power law radial surface density distribution and abrupt cut-offs), or by the best-fit centroid in the case of models with a Gaussian surface density dependence on radius. We conservatively assume our uncertainty $dR$ to be represented by half the best-fit radial width of the belt $\Delta R$ for cases where the width is well resolved, and by half the upper limit on $\Delta R$ for the three cases where the widths are unresolved (marked by the $^{\star}$ symbol in Table \ref{tab:sourcelist}). As considered later in \S\ref{sec:radunc}, this choice of $R$ and $dR$ inevitably affects our analysis, but not our main conclusions. We determine the stellar luminosity $L_{\star}$ as the integral of the observed stellar intensity across all wavelengths.

\begin{figure}
\hspace{-7mm}
\includegraphics*[width=1.12\columnwidth]{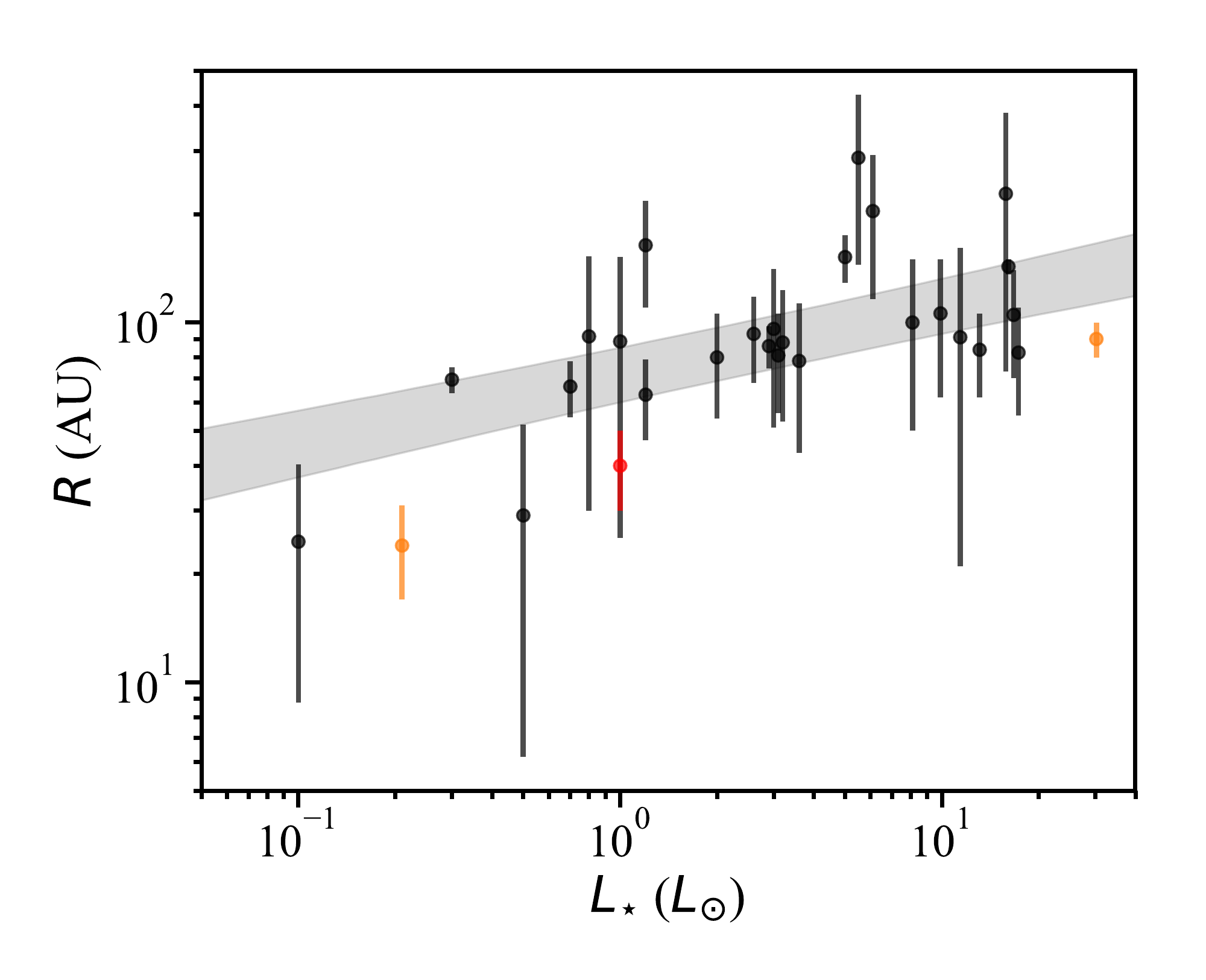}
\vspace{-8mm}
\caption{Observed planetesimal belt radii vs stellar luminosities. The black bars represent the measured extent of debris belts in case belt widths are resolved, and an upper limit to the extent in case they are unresolved (from Table \ref{tab:sourcelist}). The shaded region represent the range of power laws with likelihood within $\pm$1$\sigma$ of the best-fit, including the intrinsic scatter as well as the uncertainty on the derived parameters. The orange error bars represent the observed location of the CO snow line in the two protoplanetary discs \citep[TW~Hydrae and HD~163296,][]{Schwarz2016, Qi2013, Qi2015} and the red error bar represents the radial extent of the Kuiper belt \citep[30-50~AU,][]{Stern1997}. We assume a main-sequence luminosity of 0.16 and 34 $L_\odot$ for TW~Hydrae and HD~163296, respectively, based on their estimated stellar masses of $0.6-0.8$ and 2.3 $M_\odot$, respectively \citep{Webb1999, Natta2004}.}
\label{fig:mmlaw}
\end{figure}

\begin{figure}
\hspace{-2mm}
\centering
\includegraphics*[width=0.98\columnwidth]{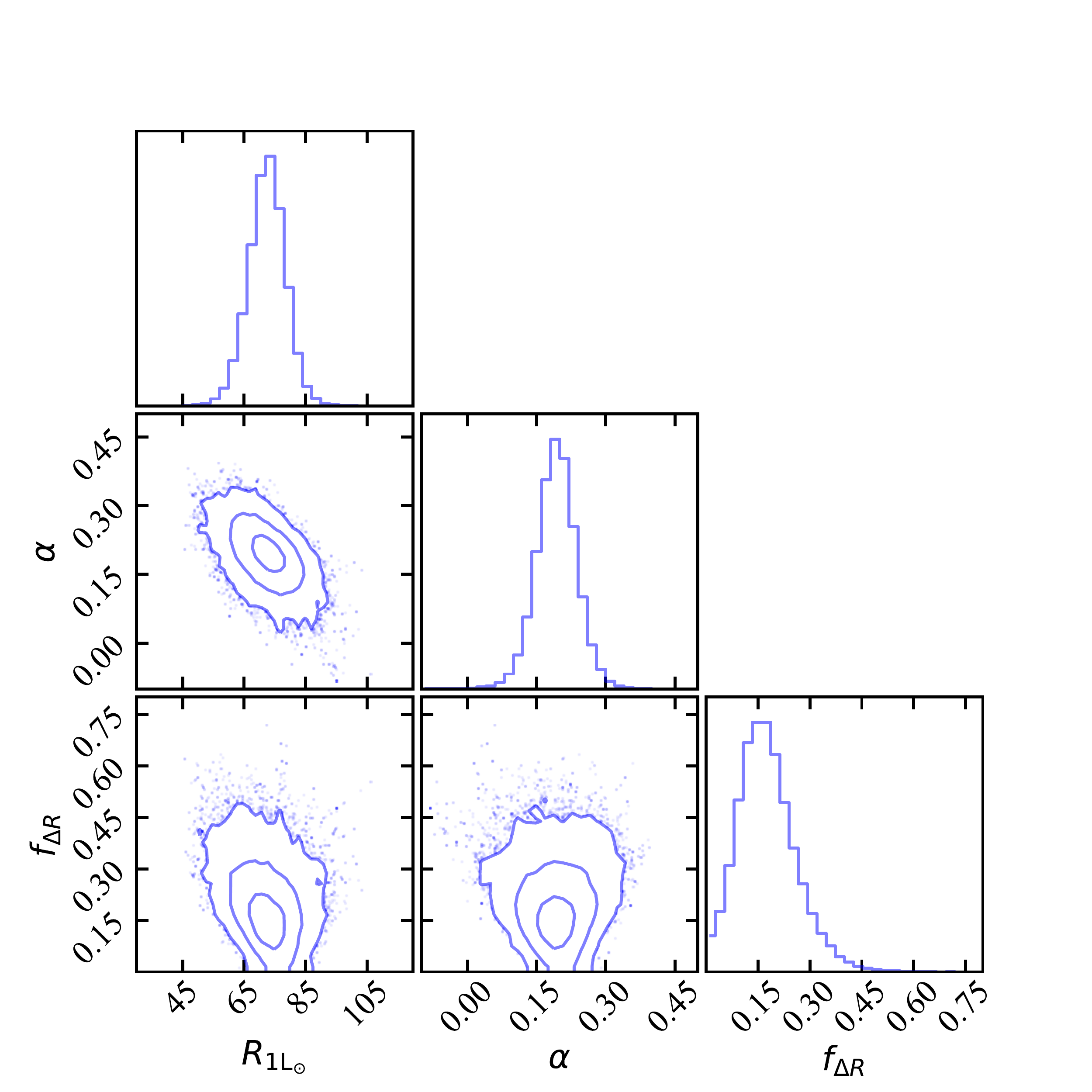}
\vspace{-3mm}
\caption{Marginalised posterior probability distributions of the power law parameters (slope $\alpha$, intercept $R_{\rm 1 L_{\odot}}$, and fractional intrinsic scatter $f_{\Delta R}$)fitted to our observed data points (Fig. \ref{fig:mmlaw}). These were sampled through MCMC methods as described in the main text. 1D histograms represent probability distributions of each parameter marginalised over the other two, whereas contour maps represent 2D probability distributions of different pairs of parameters, marginalised over the third. Contours represent the central [68.3, 95.5, 99.73] \% of the distributions. Note that this fit does not take into account observational selection effects in $[R, L_{\star}]$ space.}
\label{fig:mmlawcorner}
\end{figure}

As shown in Fig. \ref{fig:mmlaw}, we find a correlation between belt radii and the luminosity of their host star.
The correlation is well represented by a power law model where the belt locations $R_i$ (in au) are linked to their host star luminosities $L_{\star, i}$ (in $L_{\odot}$) through the form $R_i=R_{1L_{\odot}}L_{\star, i}^{\alpha}+\epsilon_i$, where $\epsilon_i$ represents the intrinsic scatter of the distribution, which we assume to follow a Gaussian distribution with standard deviation $\sigma_{\rm intr}=f_{\Delta R}R_i$. Assuming this power law model, we take an uninformative uniform prior on the free parameters $R_{1L_{\odot}}, \alpha$ and $f_{\Delta R}$, and a likelihood function described by Eq. 24 in \citet{Kelly2007}, assuming Gaussian errors on radii and taking into account the intrinsic scatter $f_{\Delta R}$. We use these to sample the posterior probability distribution of our 3 parameters through a Markov-chain Monte Carlo (MCMC) approach. We implement the latter through the \textsc{emcee} package \citep{Foreman-Mackey2013} and using the affine-invariant sampler of \cite{GoodmanWeare2010}. Taking the $50^{+34}_{-34}$ percentiles of the posterior distributions for our parameters (shown in Fig. \ref{fig:mmlawcorner}), we can set $1\sigma$ constraints on $R_{1L_{\odot}}=73^{+6}_{-6}$ au, $\alpha=0.19^{+0.04}_{-0.04}$ and $f_{\Delta R}=0.17^{+0.08}_{-0.07}$.

%particularly the inference on the magnitude of the intrinsic scatter $f_{\Delta R}$. In particular, we expect a less conservative choice for the uncertainties $dR$ (i.e., smaller uncertainties) to increase the intrinsic scatter needed to fit the data. While this choice affects the absolute measure of the intrinsic scatter, we will show in \S\ref{sec:biasalone} that it does not affect our conclusions.

\section{Bias analysis}
\label{sect:anal}

While the tight constraints on the power law parameters are indicative of a significant correlation, we need to consider whether our sample has been selected in an unbiased way within the $[R,L_{\star}]$ parameter space considered here, which we will show not to be the case. In this Section, we therefore aim to verify and quantify whether selection effects applied to an uncorrelated population could have led to the observed $R-L_{\star}$ relation.

\subsection{Selection criteria}
\label{sec:selcrit}

Three selection criteria determine whether a belt will appear on our $[R,L_{\star}]$ plot: 1) detection of excess flux due to dust at infrared (IR) wavelengths, the discovery method for planetesimal belts; 2) detection of the same excess flux at millimeter wavelengths, and 3) resolvability of the belt with currently available mm-wavelength interferometric facilities. We here describe our treatment of these effects.

\subsubsection{Infrared detectability}
For IR excess detection, we require a belt to be brighter than 3 times the typical sensitivity of \textit{Spitzer} MIPS surveys \citep[e.g.][]{Su2006} at 24 $\mu$m (taken as the largest between 0.3mJy and 2\% of the star's 24 $\mu$m flux) and 70 $\mu$m (5mJy and 5\%). If the belt is not detectable by \textit{Spitzer}, we check whether it would have been selected and detected by the \textit{Herschel} DEBRIS \citep[e.g.][]{Phillips2010} and DUNES \citep[e.g.][]{Eiroa2013} surveys at 100 $\mu$m (1.5 mJy, 5\%) and 160 $\mu$m (3.5 mJy, 5\%).
When considering detectability, if a belt of radius $R$ is spatially resolved at any wavelength, we take into account that the sensitivity to a belt's total flux density becomes different from the telescope's surface brightness sensitivity. This is because the flux density of the belt is spread over $N_{\rm res}$ resolution elements, which means that the uncertainty on the flux density becomes the telescope surface brightness sensitivity multiplied by $\sqrt{N_{\rm res}}$. We calculate $N_{\rm res}$ as the number of resolution elements covering the belt's circumference, assuming the belt is viewed face-on and its width is unresolved. Although we take this effect into account, we find that it does not have a major effect on our results in the following Sections, as only a very small fraction of belts that are detectable are also faint and/or nearby and/or large enough to not pass this selection criterion.

\subsubsection{Millimetre single-dish detectability}
For a belt to have been targeted for resolved millimeter observations, we require it to have an 850 $\mu$m flux that would have been detectable by the JCMT through the SONS survey \citep[sensitivity of $\sim$1 mJy,][]{Holland2017}. Previous millimetre detection by single dish telescopes with similar sensitivities was the main selection criterion for most of the belts in our sample (18/26), the majority of which were detected by the SONS survey itself. The remaining 8 were detected at mm wavelengths for the first time and at the same time resolved by ALMA. Of these 8, six were resolved by \citet{Lieman-Sifry2016}, who selected them to have a bright fractional 70 $\mu$m excess of at least 100, and two were resolved by \citet{Moor2017}, who selected them to be cold ($T_{\mathrm{dust}}<$140 K), high-fractional luminosity ($f>5\times10^{-4}$) belts around A stars. We use these different criteria when evaluating the bias in our sample on a star-by-star basis, but use single dish detectability when considering our stellar sample globally.

In general, while we acknowledge that the adopted telescope sensitivities may be slightly better or worse for part of the observed population, we adopt them as a close approximation to the detection bias introduced, on average, for the majority of the population of belts in the Solar neighbourhood.

\subsubsection{Millimetre interferometric resolvability}

In order to allow a radius measurement, we also require that a belt is resolvable over at least two resolution elements for the highest resolution achievable with the ALMA interferometer at the wavelength that is most sensitive to dust emission with a millimetre slope typical of nearby planetesimal belts. This corresponds to 0$\farcs$028 at 870 $\mu$m, and sets a hard lower limit on the radius of a belt that we are able to resolve.

In practice, another aspect to take into account when assessing a belt's resolvability is whether the signal to noise ratio per resolution element (or in other words, the surface brightness sensitivity) is sufficient for accurate determination of a belt's radius. In that context, we consider the fractional accuracy $dR/R$ achieved when measuring the location $R$ of a narrow ring whose width is unresolved (as we are assuming here). The uncertainty $dR$ can be estimated as $\sim \mathrm{FWHM}/\mathrm{SNR}$ where SNR is the signal to noise ratio achieved over one resolution element of size FWHM (in au) covering the ring width radially. Assuming the belt location is resolved ($R> \mathrm{FWHM}$), that the ring is face-on, and employing azimuthal averaging to boost the SNR, we can write $\mathrm{SNR}\sim F_{\nu}/(\sigma_{\mathrm{res}}\sqrt{N_{\mathrm{res}}})$, where $F_{\nu}$ is the total flux density of the belt (where $\nu\sim 345$ GHz), $\sigma_{\mathrm{res}}$ is the sensitivity per resolution element of the instrument and $N_{\mathrm{res}}$ is the number of resolution elements across the ring's circumference ($N_{\mathrm{res}}=2\pi R/\mathrm{FWHM}$). We therefore derive that $dR/R\propto \sqrt{\mathrm{FWHM}/R}\times\sigma_{\mathrm{res}}/F_{\nu}$. 

We already required that $R> \mathrm{FWHM}$ and that a belt is detectable by single-dish facilities at millimeter wavelengths ($F_{\mathrm{850 }\mu\mathrm{m}}>3$ mJy). Noting that ALMA's surface brightness sensitivity is much better than this single-dish detectability threshold ($\sigma_{\mathrm{res}}\ll F_{\mathrm{850 }\mu\mathrm{m}}$), it follows from the expression above that \textit{ALMA can accurately determine the radius of any belt that is detectable by single dish facilities.} Therefore, the only requirement we adopt for resolvability is that a belt is large enough for its diameter to be resolved over at least two resolution elements with ALMA at 870 $\mu$m ($R>0.028\arcsec$).

\subsubsection{Optical thickness of small disks}
\label{sec:optthick}
Finally, we consider whether a belt has a small enough radius and/or high enough mass for its dust emission to become optically thick (see derivations in Appendix \ref{sec:optdepder}).
The optical depth to the line of sight $\tau$ can be simply estimated for a face-on belt as the total cross sectional area in small grains divided by the on-sky area of the belt, resulting in the optical depth being proportional to the belt's fractional luminosity $f=L_{\rm dust}$$/L_{\star}$ \citep[as, for example, in][]{Jura1991, ArtymowiczClampin1997}. In particular, face-on belts with an assumed fractional width $\Delta R/R$ of 0.5 only become optically thick along the line of sight ($\tau>1$) if they have a fractional luminosity $f>2.5\times10^{-1}$.%the most massive belt in our observed sample has a total grain cross-sectional area of $\sim$88 AU$^2$; assuming a fractional width $\Delta R/R$ of 0.5, a belt this massive would reach a face-on optical thickness $\gtrsim$1 only for radii $\lesssim5.6$ AU.
We also consider an edge-on geometry, assuming a uniform density ring with $\Delta R/R$ of 0.5 and a scale height $H/R$ of 0.1. In this case, their maximum optical depth along the line of sight reaches values $>1$ for fractional luminosities $f>7.1\times10^{-3}$. Since only few of the most massive belts that we consider in the following subsections (and only one of our observed belts) are affected, this effect is largely negligible for our population study.

\subsection{Understanding the bias in $[R,L_{\star}]$ space}
\label{sec:undersbias}

We here test the hypothesis that these selection effects alone applied to a population uncorrelated in $[R,L_{\star}]$ space could reproduce our data. We use a Monte Carlo approach, drawing a large population of model belts uniformly in log$_{10}$$([R,L_{\star}])$ space and passing them through our selection criteria (\S\ref{sec:selcrit}). However, assessing detectability and resolvability requires a model connecting a belt's $[R,L_{\star}]$ to its belt and host star's flux as observed from Earth at several wavelengths. We derive the host star's flux at a given wavelength assuming blackbody emission, and deriving all other stellar properties
 from $L_{\star}$ assuming it has reached the main sequence, interpolating tabulated values of \citet{PecautMamajek2013}\footnote{\url{http://www.pas.rochester.edu/~emamajek/EEM_dwarf_UBVIJHK_colors_Teff.txt}}.

To derive a disk's flux from $[R,L_{\star}]$ we use a simple, narrow belt model as described in \citet{Wyatt2008}, whose SED is described by a modified blackbody characterised by a temperature $T$, a fractional luminosity $f$, and a flux density ($F_{\nu}$) falling off as a power law with slope $(-2-\beta)$ at wavelengths larger then a given $\lambda_0$. We remind the reader that a blackbody grain of temperature $T$ derived from the SED would lie at a distance from the star equal to $R_{\rm BB}$, the so-called blackbody radius \citep[see Eq. 3 in][]{Wyatt2008}. In practice, small grains dominating the SED are always hotter than blackbody, meaning that the true radius $R$ of a belt as determined by resolved mm-wavelength observations is always greater or equal to $R_{\rm BB}$. Throughout this work, we will use $R_{\rm BB}$ as an equivalent measure of temperature in order to calculate the belt flux. Thus, calculating the flux of a belt of known $[R,L_{\star}]$ requires introducing extra free parameters $R_{\rm BB}/R$, $f$, $\lambda_0$, $\beta$, as well as $d$, the distance to the star from Earth. 

\begin{figure}
\vspace{-0mm}
 \hspace{-2mm}
  \includegraphics*[scale=0.42]{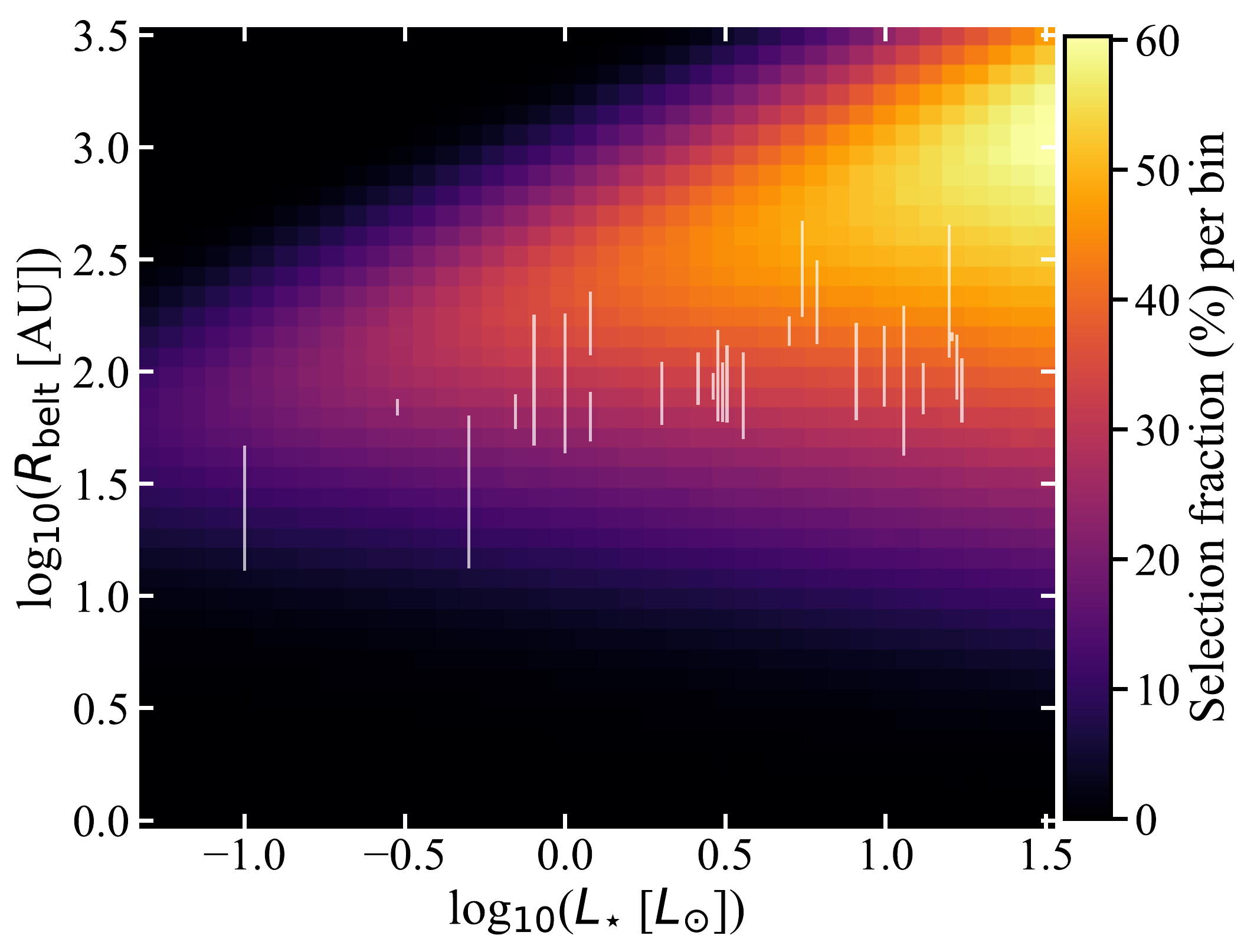}
\vspace{-6mm}
\caption{Selection probability (\%) per $\rm log_{10}$$([R,L_{\star}])$ bin of a simulated population of belts drawn assuming a log-uniform distribution of fractional luminosities $f$ between $[10^{-7},10^{-2}]$, a uniform distribution of $R/R_{\rm BB}$ between $[1.2,5.5]$, of $\lambda_0$ between $[29.4, 592.0]$ $\mu$m, and of $\beta$ between $[0.2, 1.9]$. Belt distances from Earth are drawn from an isotropic distribution ($N(d)\propto d^2$) out to 150 pc. White vertical bars represent our sample of belts currently resolved at millimeter wavelengths from Fig. \ref{fig:mmlaw}.
} 
\label{fig:detectmap}
\end{figure}

This means that we have to make assumptions for these parameters that will impact the detectability of belts and hence affect the observational bias. We will test the effect of changing these assumptions in Appendix \ref{sec:testass}, and here show results for our `fiducial' model. For the latter, we assume a prior log-uniform distribution for $f$, a linearly uniform distribution for $R/R_{\rm BB}$, $\lambda_0$ and $\beta$, and log-uniform distributions for $R$ and $L_{\star}$ which are not correlated with one another.
The boundaries of the distributions of $\mathrm{log}_{10}$$([R,L_{\star}])$ are the same as the plot boundaries in Fig. \ref{fig:detectmap}. For the other parameters, we resort to empirical evidence from the extremes within our resolved belt sample to set our prior boundaries for $f$ between $[10^{-7},10^{-2}]$, for $R/R_{\rm BB}$ between $[1.2,5.5]$, for $\lambda_0$ between $[29.4, 592.0]$ $\mu$m, and for $\beta$ between $[0.2, 1.9]$. Note that we will refer to this as a `static' model, as (at least initially) we do not consider a belt's evolution with time and its effect on these observables.

For each of the $L_{\star}$ columns in Fig. \ref{fig:detectmap} we synthesize a population of $4\times10^5$ belts, $10^4$ for each radius $R$ sampled in the vertical direction. Each of these belts is then assigned a set of parameters $[f, R/R_{\rm BB}, \lambda_0, \beta]$ drawn from the assumed distributions described in the previous paragraph, and a distance from Earth $d$ drawn assuming a spherically isotropic distribution of stars ($N(d)\propto d^2$) out to a distance of 150pc, which is approximately the distance to the furthest star in our observed sample.
Then, Fig. \ref{fig:detectmap} displays the fraction of the population of $10^4$ belts simulated in each log$_{10}[R, L_{\star}]$ bin that would pass our selection criteria derived in \S\ref{sec:selcrit}.

We find that the region where belt radii would have been selected has a triangular shape in $[R, L_{\star}]$ space. The upper and lower limits to selected radii are dominated, respectively, by the disk's detectability at 70 and 850 $\mu$m. This is because at any given stellar luminosity $L_{\star}$, for a fixed fractional luminosity $f$, belts increasingly further from the star quickly become too cold for 70 $\mu$m detection (due to the steep short-wavelength slope of the blackbody function). On the other hand, once again for a fixed fractional luminosity $f$, belts increasingly closer to the star more slowly become too warm for 850 $\mu$m detection (due to the shallower long-wavelength slope of the modified blackbody function). We remind the reader that the fact that belts can become too warm for sub-millimeter detection is because for a constant fractional luminosity, as assumed here, the dust mass is not constant but increases with radius ($M_{\rm dust}\propto R^2$), hence decreasing with temperature.

We highlight the fact that the colour map of Fig. \ref{fig:detectmap} shows how the selection probability per bin varies in $[R, L_{\star}]$ space; this significantly differs from the number of selected stars per bin, which we present and discuss in Appendix \ref{sec:resbeltab}. Therefore, the colour map in Fig. \ref{fig:detectmap} should not be compared with the density of observed points. Rather, we are interested in how vertical cuts in the colour map at a given $L_{\star}$ compare with the radius $R$ of our observed belts.

\subsection{Can the $R-L_{\star}$ relation be explained by selection bias alone?}
\label{sec:biasalone}

\begin{figure*}
 \centering
 \hspace{-3mm}
  \includegraphics*[scale=0.6]{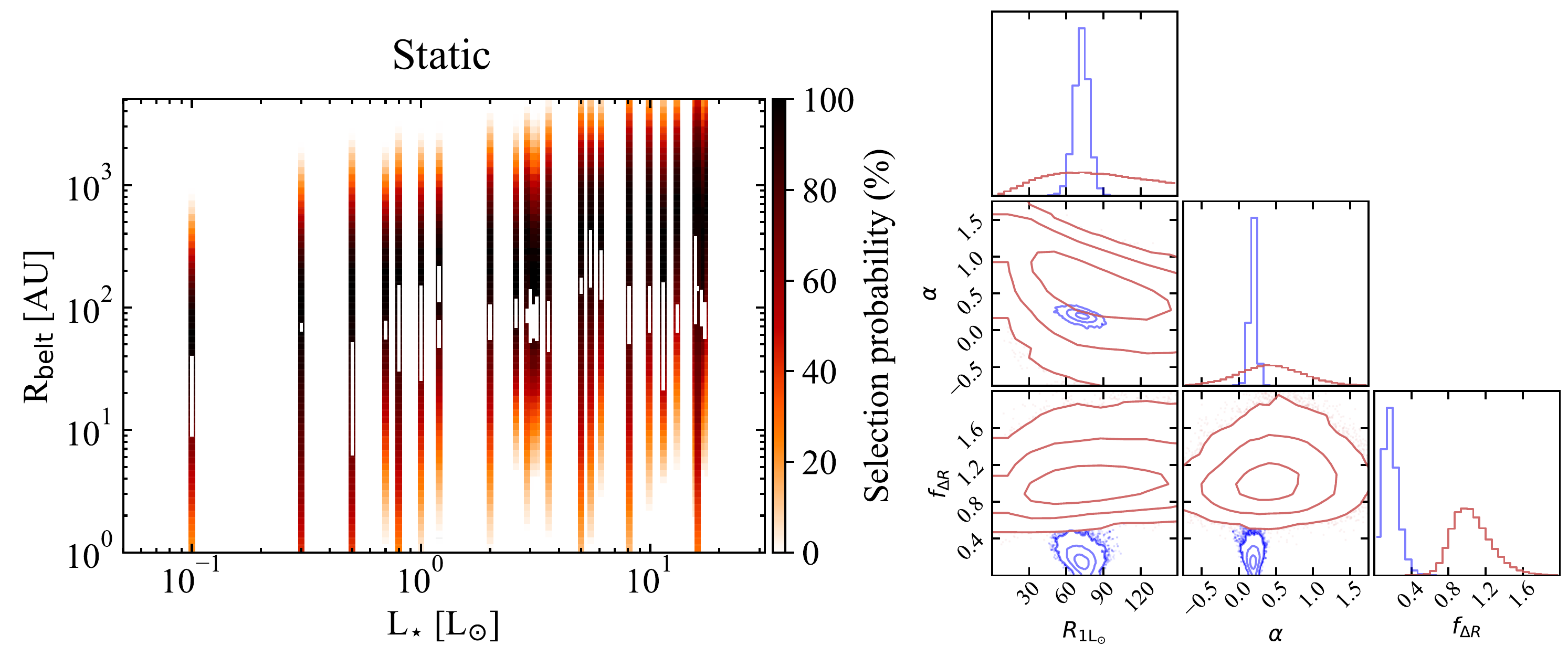}
\vspace{-5mm}
\caption{Results of Monte Carlo simulations of static model belt populations. \textit{Left:} Vertical color strips represent the normalized selection probability of belts in $\rm log_{10}$$(R)$ space for each of the stars in our sample, given their luminosity $L_{\star}$ and distance from Earth $d$. White vertical bars represent our sample of belts currently resolved at millimeter wavelengths from Fig. \ref{fig:mmlaw}. \textit{Right:} Marginalised probability distributions analogous to Fig. \ref{fig:mmlawcorner}, showing the results from fitting the data in blue). For comparison, red histograms and contours represent marginalised probabilities of randomly drawing a given set of parameters from an uncorrelated population of model belts, after accounting for observational selection effects%. 1D histograms represent probability distributions of each parameter marginalised over the other two, whereas contour maps represent 2D probability distributions of different pairs of parameters, marginalised over the third. Contours represent the central [68.3, 95.5, 99.73] \% of the distribution. Blue solid lines represent marginalised posterior probability distributions of the parameters given our observed sample, and should be compared with the model distributions.
} 
\label{fig:corner_uncorr_static_fiducial}
\end{figure*}

The question we aim to answer is `Given our observed population of 26 stars, using our simple belt model with its assumptions and taking into account selection effects, what is the probability of having found our $R-L_{\star}$ correlation if no correlation was present?'.
For each star in our sample of 26 resolved belts, we therefore take its known luminosity $L_{\star}$ and distance $d$ from Earth and create a population of $10^{6}$ belts, with the same fiducial model assumptions as employed in \S\ref{sec:undersbias}. For each star, we evaluate the fraction of belts that would be selected as a function of radius following our selection criteria. This yields a selection probability distribution of radii for each of our 26 stars (vertical color strips in Fig. \ref{fig:corner_uncorr_static_fiducial}, left).

From each star's probability distribution, we draw a single radius and fit a power law model to the simulated $R-L_{\star}$ dependence through MCMC fitting, as done for the observed data (\S\ref{sect:res}). Using this approach requires assigning an uncertainty $dR$ to the simulated radii $R$ drawn for each $L_{\star}$. As this uncertainty affects the derived intrinsic scatter of the relation (\S\ref{sect:res}), given that we want to ensure fair comparison between the scatter of the simulated and observed populations, we assume each drawn radius at a given stellar luminosity to have the same fractional uncertainty $dR/R$ as that of the corresponding observed belt.

We repeat this MCMC fitting for 10$^5$ simulations of the $[R-L_{\star}]$ relation, and each time retrieve the set of best-fit parameters $R_{1L_{\odot}}, \alpha$ and $f_{\Delta R}$. This allows us to obtain a 3D probability distribution of drawing the 3 observed power-law parameters from an uncorrelated $[R,L_{\star}]$ population, which we show in Fig. \ref{fig:corner_uncorr_static_fiducial}, right. These simulated probability distributions (shown in red) can then be compared with the posterior probability distributions of the 3 parameters inferred from our observed data (blue, as derived in \S\ref{sect:res}). 

The probability distributions for our fiducial model in Fig. \ref{fig:corner_uncorr_static_fiducial} indicate that there is a modest probability of drawing a power law slope and intercept similar to the ones observed. Both the increasing upper envelope of IR detectability and the fact that more luminous stars in our sample tend to lie at larger distances $d$ from Earth (increasing their smallest detectable radius) contribute to the result. On the other hand, we find that the marginalised probability (over all slopes and intercepts) of finding an intrinsic scatter $f_{\Delta R}$ within $\pm$1$\sigma$ of our observed value ($0.17^{+0.08}_{-0.07}$) is below our capability to sample ($<10^{-5}$). In other words, none of our $10^5$ simulated $[R-L_{\star}]$ relations displays an intrinsic scatter within $\pm$1$\sigma$ of our observed median value. This indicates that randomly drawing a highly correlated dataset such as ours from an uncorrelated population after taking biases into account is very unlikely. This is mainly driven by the spread of our observed data points about the best-fit power law being much smaller than we would obtain from an underlying uncorrelated population.
%In turn, this large scatter of the simulated uncorrelated population is dominated by the width of the region between the maximum and minimum detectable radii.%, which is mostly determined by the upper boundary of the fractional luminosity distribution (see \S\ref{sec:testass}).

Of course, this conclusion is dependent upon our assumptions for the set of parameters [$f$, $R/R_{\rm BB}$, $\lambda_0$, $\beta$] characterising the belt population. In Appendix \ref{sec:testass}, we examine the effect that changing each of these parameters has on our conclusion above. 
%In summary, these tests show that a comprehensive belt population model with a distribution of parameters [$f$, $R/R_{\rm BB}$, $\lambda_0$, $\beta$] as well as $[R, L_{\star}]$ is needed to understand whether the observed $R-L_{\star}$ may be explained by observation bias alone. This is highly degenerate problem, particularly if we consider our relatively small population of 26 observed belts.
%Instead of attempting a multi-dimensional fit, in this first attempt, we simply tried to reproduce our $R-L_{\star}$ relation through sensible choices of free parameters informed by our observed sample and previous IR population studies.
%instead of trying to reproduce the observed population distribution of all these parameters at the same time, which is likely a highly degenerate problem given our population of only 26 observed belts, we here simply attempted to reproduce our $R-L_{\star}$ relation through sensible choices of free parameters informed by our observed sample and previous IR population studies.
In summary, we find that while we cannot fully rule out that a specific combination of parameter assumptions may explain the observed $R-L_{\star}$ relation, none of our reasonable sets of assumptions (informed by our observed sample and previous IR population studies) can reproduce the observed population.
In particular, the formal probability of drawing a relation consistent with ours from an uncorrelated underlying population remains exceedingly low for all our tested assumptions, even for model populations with $R/R_{\rm BB}$, $\lambda_0$ and $\beta$ fixed to a constant value rather than drawn randomly from a range of values. This is mainly driven, in all cases, by the observed scatter being much lower than predicted for an underlying uncorrelated $[R, L_{\star}]$ population, which indicates that a true $R-L_{\star}$ relation in the underlying population of belt radii is likely necessary to explain our observed trend.

\subsection{On the definition of the radius uncertainty $dR$}
\label{sec:radunc}
We note that our choice of uncertainties $dR$ on the observed radii being equal to half the belt widths affects the derived parameters and their uncertainties. Nonetheless, we made this choice in light of two fundamental issues.
First, true uncertainties on the belt radius and width, which should be independently quantifiable, are not measured in a consistent way in different literature works. The difficulty lies in the problem that most belts were fitted independently using a variety of parametrizations (see caption of Table \ref{tab:sourcelist}), resulting in parameter uncertainties that do not easily translate to a radius uncertainty $dR$. 
Second, the choice of radius is itself dependent on which part of a belt is most relevant for its formation, which depends on which theory we are trying to test (see discussion in \S\ref{sect:disc}). %The choice of radius then also affects its own uncertainty, as well as our derived power law parameters.
Although our choice of $R$ as the `middle' radius is somewhat arbitrary, we deem it a more robust representation of where most of the dust is located than, for example, an inner or outer radius. 
The choice of radius of course also affects its own uncertainty, as well as our derived power law parameters.
A major effort in consistently reanalyzing all archival datasets, which is beyond the scope of this paper, would be needed to enable us to change the definition of radius and measure its associated uncertainty.

Our main conclusion on the significance of the correlation stems from the fact that the scatter in the observed radii is small, and in particular smaller than would be expected from an underlying uncorrelated population. In other words, measured radii don't fill the detectable $[R,L_{\star}]$ space as well as expected from an uncorrelated model population. This is despite the conservatively large uncertainties $dR$ that we assumed. Then, assuming smaller uncertainties would increase the inconsistency of the data with the model expectations, since observations would fill even less of the $[R,L_{\star}]$ parameter space.

Investigating this issue more carefully, we can compare the intrinsic scatter $f_{\Delta R}$ of our measurement with that expected from a randomly drawn, uncorrelated model belt population, as done in \S\ref{sec:biasalone} above. This time, we test the effect of our assumption on $dR$ by recalculating probability distributions after fixing $dR/R$=0.1 for both observed and simulated belt populations, for any stellar luminosity. These are shown as dotted lines in Fig. \ref{fig:corner_uncorr_testass_1}, top right, where the top row of Fig. \ref{fig:corner_uncorr_testass_1} is otherwise equivalent to Fig. \ref{fig:corner_uncorr_static_fiducial}. As expected, this less conservative choice for the uncertainties $dR$ (i.e., smaller uncertainties) increases the intrinsic scatter needed to fit the data. However, the same change applies to the model population, leaving the comparison between the two, and therefore our conclusion on the existence of an underlying $R-L_{\star}$ relation, unaffected. Practically, this is because the observed scatter of the data and the `observed' scatter of the model population result from a combination of the intrinsic scatter and the assumed uncertainties $dR$. Then, changing the uncertainties in the same way for both the data and the model will only cause the derived intrinsic scatter to compensate in the same way for both, making the comparison largely independent of the choice of uncertainties $dR$.

%the main result is an increase in the derived intrinsic scatter $f_{\Delta R}$ for both the observed and simulated data. Crucially, however, the probability distributions for the intrinsic scatter increase by a similar amount for both the observed and simulated data, proving that the comparison between the two is largely independent of our assumption for the radius uncertainty.  
%Logically, a less conservative choice for the uncertainties $dR$ (i.e., smaller uncertainties) increases the intrinsic scatter needed to fit the data. However, as shown by our test in \S\ref{sec:biasalone}, the same change applies to the model population, leaving the comparison between the two, and our conclusion on the existence of an underlying $R-L_{\star}$ relation, unaffected.

\subsection{Quantifying the effect of selection bias on the uncertainty on derived power law parameters}
\label{sec:corr}

Having concluded that an underlying correlation between belt radii and their host star's luminosity is likely necessary to explain the data, we here aim to quantify how selection bias affects the $[R_{1 L_{\odot}}, \alpha, f_{\Delta R}]$ parameters derived through our power law fit to the $R-L_{\star}$ relation in \S\ref{sect:res}, and in particular their uncertainties.
We adopt the same MCMC fitting approach as in \S\ref{sect:res}, but this time we modify the likelihood function of the power law parameters given the data to include selection effects, following the Bayesian method described in \S5 of \citet{Kelly2007}. In summary, this acts by assigning higher probabilities to belts that are harder to detect, by weighting the contribution of the likelihood function from each belt radius by the inverse of the selection probability at that radius, as derived above in \S\ref{sect:anal}. This effectively counterbalances our selection effects and debiases our inference on the model parameters. Of course, our debiasing method remains dependent on the same assumptions for [$f$, $R/R_\mathrm{BB}$, $\lambda_0$ and $\beta$] as considered in the previous subsections.

We here make the assumption of a belt population with log-uniform fractional luminosity and with fixed $R/R_\mathrm{BB}$, $\lambda_0$ and $\beta$. Note that as demonstrated in Appendix \ref{sec:testass}, fixing these values rather than drawing them from a distribution does not change the result significantly compared to the fiducial model. 
We find $R_{1 L_{\odot}}=66.8^{+7.7}_{-11.8}$, $\alpha=0.19^{+0.05}_{-0.06}$ and $f_{\Delta R}=0.23^{+0.27}_{-0.10}$, where these new debiased parameters are consistent with the biased ones. As expected, the uncertainties on the derived parameters increased because this debiased fitting takes into account that some undetected belts may lie in regions of low selection fraction. These debiased parameters represent the properties of the underlying population after taking biases into account. Therefore, the fact that these parameters are inconsistent with the expectation of an uncorrelated population (e.g. $\alpha=0$ and large $f_{\Delta R}$), and that they are well constrained within their uncertainties confirms that the radius-luminosity relation and the derived slope are robust against observational biases, at least for the fiducial population assumptions considered here.

\section{Discussion}
\label{sect:disc}
Throughout \S\ref{sect:anal}, we analysed the effect of observational biases on the belt radius - stellar luminosity relation and demonstrated that it is likely that the observed relation is caused by a true underlying correlation between the two parameters.
We now analyse what the origin of this $R-L_{\star}$ relation may be and whether it could constrain the belts' formation location within protoplanetary disks. In order to do that, we need to consider the effect of the collisional evolution over the belts' lifetime.

\subsection{Steady-state collisional evolution} 
\label{sec:steadystateevol}

\begin{figure}
\vspace{-0mm}
 \hspace{-2mm}
  \includegraphics*[scale=0.42]{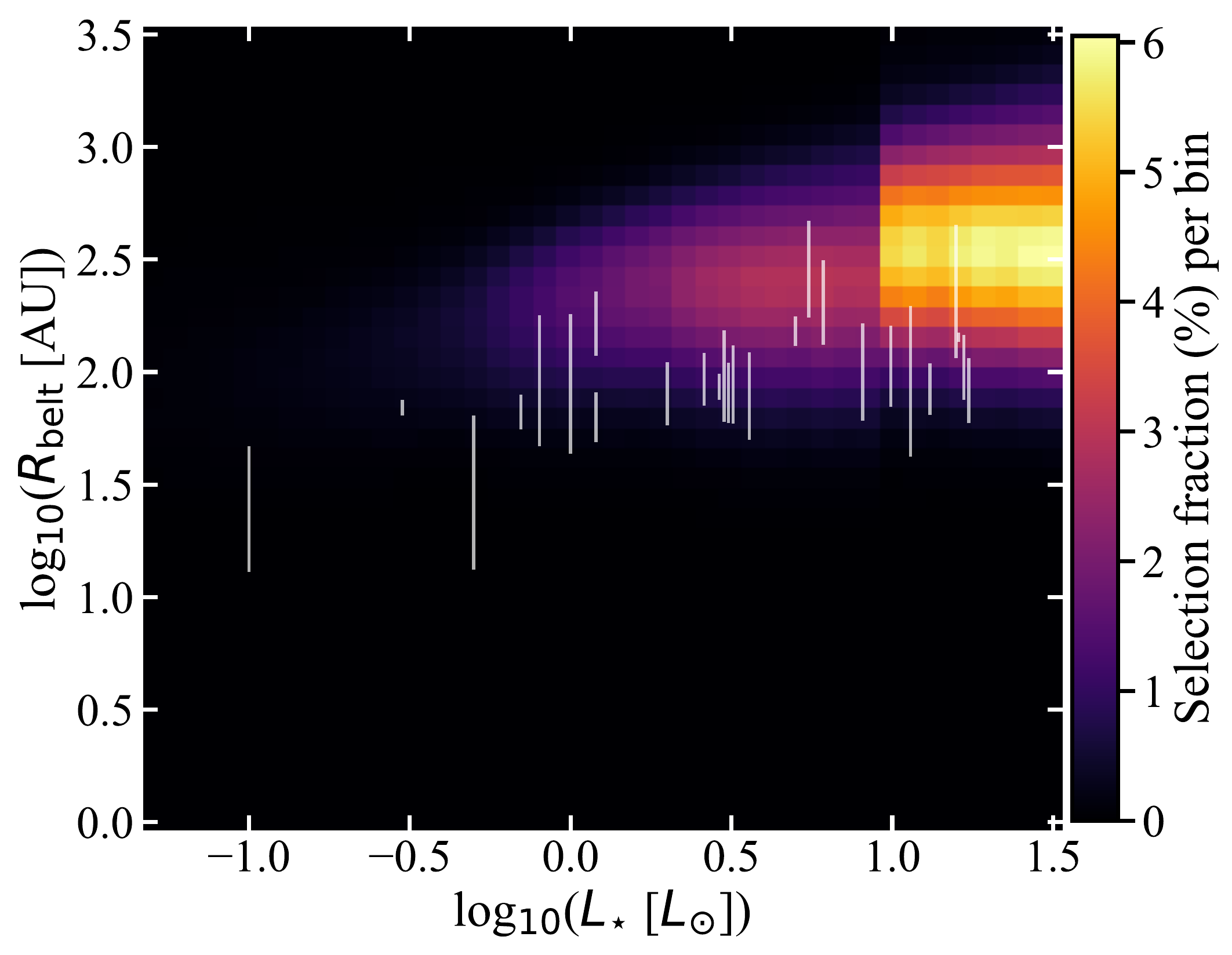}
\vspace{-6mm}
\caption{Selection probability (\%) per $\rm log_{10}$$([R,L_{\star}])$ bin of a simulated population of belts whose dust mass has been evolved according to the model and best-fit parameters of \citet{Wyatt2007b} and \citet{Sibthorpe2018}. For a star of a given luminosity, an age is drawn from a uniform distribution up to the smallest between its main-sequence lifetime and the age of the universe. Belt distances from Earth are randomly drawn from an isotropic distribution ($N(d)\propto d^2$) out to 150 pc. White vertical bars represent our sample of belts currently resolved at millimeter wavelengths from Fig. \ref{fig:mmlaw}.} 
\label{fig:detectmap_evolved}
\end{figure}

A clear outcome of our bias analysis in \S\ref{sect:anal} was that, regardless of the assumptions in our model, the simulated populations after considering observational bias showed a scatter in radii that is much larger than observed. Under the log-uniform fractional luminosity assumption, the model prediction is that a large number of belts should have been detected and resolved at larger and smaller radii than the observed sample (as shown in Fig. \ref{fig:corner_uncorr_static_fiducial}, left). On the other hand, models with a log-uniform distribution of belt mass (see Appendix \ref{sec:testass} and Fig. \ref{fig:corner_uncorr_testass_1}, bottom row) do a significantly better job of reproducing the lack of radii much larger than observed, but does a significantly worse job at reproducing the lack of belts with radii much smaller than observed. Overall, the distribution of dust masses (or fractional luminosities) is the parameter that most significantly affects the scatter of the simulated belt populations.

What our static model of \S\ref{sect:anal} did not consider is that belts are known to deplete and grind down over time, causing a decrease of their mass and fractional luminosity \citep[e.g.][]{Spangler2001}. This decrease is faster for belts that have smaller radii and that have a higher mass stellar host, due to their planetesimals colliding at higher velocities. Therefore, for the same initial belt mass and radius at the beginning of collisional evolution, if we let belts around different stars evolve to the same system age, belts around low-luminosity stars and further from the star will be more massive than belts around higher luminosity stars and closer to the star. This implies that at a given system age, belts closer to the star and around more luminous stars will be less detectable. Conversely, we also need to consider that more luminous stars have a shorter main-sequence lifetime and are therefore on average observed at a younger age.

To test these effects expected from collisional evolution, we once again resort to Monte Carlo methods and simulate the belt population predicted by the steady state collisional evolution model described in \citet{Wyatt2008}. We assume that belts initiate collisional evolution within protoplanetary disks, and therefore that they have been collisionally evolving for the entire lifetime of the star. The evolution of belt mass according to this model is almost flat up to an age roughly corresponding to the collision timescale of the largest bodies within the belt, after which the mass decreases with time $t$ following $1/t$. 

This steady state collisional cascade model fits the observed evolution of IR excesses around both A and FGK stars \citep{Wyatt2007b, Kains2011, Sibthorpe2018}, given some reasonable assumptions and other fitted parameters, which were found to differ for the two spectral type categories. We here adopt exactly the same assumptions and best-fit parameters to examine the effect collisional evolution has on the observed belt population in $[R, L_{\star}]$ space. In particular, for both spectral type categories, the model assumes a universal belt fractional width of $\Delta R/R=0.5$, a grain density typical of silicates ($\rho$=2700 kg m$^{-3}$), a proper eccentricity of $e=0.05$, an initial blackbody radius distribution of the belt population following $N(R_{\rm BB})\propto R_{\rm BB}^{\gamma}$, and an initial belt mass that follows a log-normal distribution with a fitted centroid $M_{\rm mid}$ and a standard deviation of 1.14 dex.

The distributions of both radii and initial masses are independent of stellar properties within each of the two spectral categories. For A (FGK) stars, fitted parameters with their best-fit values were $\gamma=-0.8$ ($\gamma=-1.7$), $M_{\rm mid}=10$ M$_{\oplus}$ ($M_{\rm mid}=2.1$ M$_{\oplus}$), with the maximum planetesimal size $D_{\rm c}=60$ km ($D_{\rm c}=450$ km) and dispersal threshold planetesimal strength $Q_{\rm D}^{\ast}=150$ J kg$^{-1}$ ($Q_{\rm D}^{\ast}=500$ J kg$^{-1}$) setting the time evolution. As no constraints are present to date for M stars, we assume the same parameters as for the FGK population. 
Finally, since the model works by evolving belts located at their blackbody radii (which is consistent with the fact that the model was fitted to blackbody radii rather than true radii at IR wavelengths), we still need to make an assumption for the $R/R_{BB}$ distribution of the population. As in our fiducial model of \S\ref{sect:anal}, we assume a uniform distribution of $R/R_{BB}$ between the minimum and maximum of our observed belt population.

\begin{figure*}
 \centering
  \hspace{-3mm}
  \includegraphics*[scale=0.6]{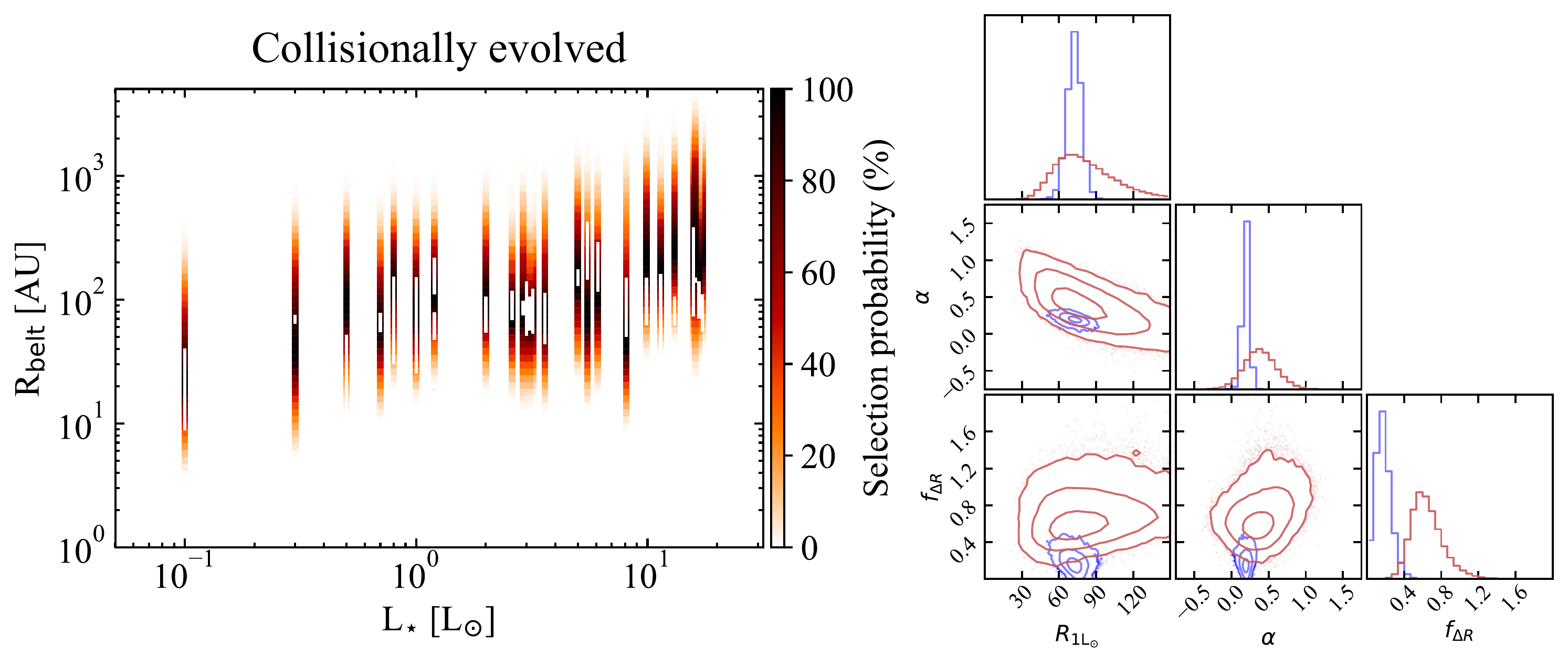}
\vspace{-7mm}
\caption{
Results of Monte Carlo simulations of model belt populations, whose mass has been evolved according to the model and best-fit parameters of \citet{Wyatt2007b} and \citet{Sibthorpe2018}. For both columns, lines and symbols have the same meaning as in Fig. \ref{fig:corner_uncorr_static_fiducial}.
} 
\label{fig:corner_uncorr_evolved_rbb}
\end{figure*}

Informed by this collisional evolution model, we once again simulate a population of 10$^6$ belts with the distribution of initial masses and radii that best fits the IR population. 
We evolve belts around each star to a random age drawn from a uniform distribution up to the lowest between the star's main sequence lifetime and the age of the universe. In Fig. \ref{fig:detectmap_evolved} we show the selection fraction per $[R, L_{\star}]$ bin (analogously to Fig. \ref{fig:detectmap}), assuming an isotropic stellar population out to a distance of 150 pc. 

The main difference between Fig. \ref{fig:detectmap_evolved} and Fig. \ref{fig:detectmap} is that a new lower envelope of detectability appears at a larger radius than before, as belts that are closer to the star evolve faster and have their mass and hence flux dropping below detectability at any given age. 
If all belts were evolved to the same age, the dependence of this lower envelope on the stellar luminosity would be $R\propto L_{\star}^{0.12}$ \citep[combining Eq. 6, 14, 15, and 16 from][and taking the approximation $L_{\star}\propto M_{\star}^4$]{Wyatt2008}. However, including the effect that less luminous stars are, on average, older than more luminous ones causes this lower envelope of detectability to be nearly flat. At the same time, this effect produces a steep dropoff in the number of detectable belts around stars of increasingly lower luminosity, as most of these belts have evolved for longer and hence depleted below detectability.
The sharp discontinuity in the color map at high luminosities is caused by the difference in the best-fit parameters fitted to the A and FGK star population at IR wavelengths, which suggests that A stars evolve at a faster rate, but also start with more massive belts. %Evolving the belt population to ages older than the 100 Myr shown here causes this lower envelope to move up, as belts with larger radii will have ground down for a long enough time to become undetectable.

As mentioned for Fig. \ref{fig:detectmap} in \S\ref{sec:undersbias}, we underline that the colour map in Fig. \ref{fig:detectmap_evolved} does not consider the luminosity function $N(L_{\star})$ in the Solar neighbourhood and therefore should not be interpreted as the number of stars in $[R, L_{\star}]$ space, which we show and discuss in Appendix \ref{sec:resbeltab}. Once again, this is because we are not interested in reproducing the population density in $[R, L_{\star}]$ space, but the our observed $R(L_{\star})$ relation given our sample of stars, with their luminosities, masses and ages. 

We therefore proceed to quantify whether this steady state collisional evolution model for planetesimal belts can explain our observed trend as in \S\ref{sec:biasalone}, by quantifying the selection probability for each of the stars in our sample (see Fig. \ref{fig:corner_uncorr_evolved_rbb}, left), given their $L_{\star}$ and distance to Earth $d$. We evolve their belt mass to their observed age (choosing best-fit values reported in the literature, see Table \ref{tab:sourcelist}), taking into account the distribution of belt radii from the collisional evolution model ($N(R_{\rm BB})\propto R_{\rm BB}^{\gamma}$). We then sample each of these probability distributions 10$^5$ times and calculate the slope, intercept and intrinsic scatter of the simulated $R-L_{\star}$ relations. The simulated probability distributions of the 3 power law parameters are shown in Fig. \ref{fig:corner_uncorr_evolved_rbb} (right, red), where they can once again be compared to the probability distributions derived from the data (blue). 

We find that the steady state collisional evolution applied to a population of belt radii that is not correlated with their host star's luminosity is likely to produce a $R-L_{\star}$ relation with a slope and intercept close to those shown by the data. Compared to our static belt model, the probability of drawing a dataset with an intrinsic scatter within $\pm1\sigma$ of that observed (for any slope and intercept) increases significantly from the $<10^{-5}$ derived from Fig. \ref{fig:corner_uncorr_static_fiducial} (right) to $2.6\times10^{-3}$. This confirms the qualitative result of Fig. \ref{fig:corner_uncorr_evolved_rbb} (left), showing that collisional evolution coupled to observational bias can reproduce the observed $R-L_{\star}$ relation much better than a static model (Fig. \ref{fig:corner_uncorr_static_fiducial}, left).
Despite the improvement, however, the probability of drawing an intrinsic scatter as low as that of the observed population remains quite low. If we formally consider the chance of drawing, at the same time, a slope, intercept and intrinsic scatter within $\pm1\sigma$ of the observed values, the probability drops to an even lower value of $10^{-4}$. 

%Our simulations show that the observed $R-L_{\star}$ relation, particularly its power law slope and intercept, can be qualitatively explained by the collisional evolution of a population of belts with radii that are not correlated with their host star's luminosities. Nonetheless, the low intrinsic scatter observed remains difficult to explain even with such an evolving model population. 
This indicates that one or more of the assumptions of the evolutionary model may not accurately describe the observed population. For example, the radii at which planetesimal belts form may not be well represented by a simple power law distribution as a function of blackbody radius ($N(R_{\rm BB})\propto R_{\rm BB}^{\gamma}$), as the comparison between the data and our simulations suggests that belts may not form as far out and/or as close in as we could have detected them. A larger sample of resolved belts and a simultaneous fit of the collisional evolution model to both the population of resolved radii and IR excesses as a function of age is necessary to establish whether different combinations of model parameters may quantitatively reproduce the observed low scatter of the $R-L_{\star}$ relation.

\subsection{A preferential formation location for planetesimal belts in protoplanetary disks?}
\label{sec:prefformloc}

An alternative explanation for the low scatter observed in the belt radii is that planetesimal belt locations could be clustered at radii that depend on their host star's luminosity. This would indicate a preferential location for planetesimal belt formation in protoplanetary disks. This hypothesis is further supported by the location of the Edgeworth-Kuiper belt in our own Solar System \citep[$\sim$30-50 au,][]{Stern1997} being close to the expectation from the $R-L_{\star}$ relation seen in Fig. \ref{fig:mmlaw}, especially when considering that it does not suffer from the observational biases discussed in this work. 

The question then arises as to what could cause planetesimal belts to form at a specific range of radii that correlate with the host star's luminosity. 
As mentioned in \S\ref{sect:intro}, planetesimal belt formation requires grain growth to lead to the formation of planetesimals, but also a mechanism to either stop these planetesimals growing further to form planets, or to grow them into planets rapidly enough that several generations of planetesimals may be produced. 
Below, we consider possible scenarios that may fulfil these requirements for planetesimal belt formation.

\subsubsection{Planetesimal formation and the CO snow line}
\label{sect:planetesimalform}

It is now well established that formation of planetesimals from $\mu$m-sized interstellar grains requires overcoming several growth barriers that are dictated by collisional physics and the interaction between solids and gas in protoplanetary disks. Collisional bouncing, fragmentation and erosion all act to slow the growth timescale of solids to the point they are lost to the star via radial drift before they can grow any further \citep[for a review, see][and references therein]{Birnstiel2016}. A promising way to overcome these barriers is through particle overdensities leading to gravitational collapse, where such concentrations in the forms of disk substructure have recently started being discovered through high-resolution dust imaging of protoplanetary disks \citep[e.g.][]{vanderMarel2013, Casassus2013, Marino2015mwc758, ALMAPartn2015, Andrews2016, Isella2016, Loomis2017, Fedele2017}. 

These overdensities can be caused by different physical mechanisms; we direct the reader to \citet{Johansen2015} for a review. We here focus on the CO snow line and its role in planetesimal formation; this is motivated by the fact that the radial location of the only two observationally-inferred CO snow lines \citep{Qi2013, Qi2015, Schwarz2016} lies close to our $[R-L_{\star}]$ relation (Fig. \ref{fig:mmlaw}). It has been theoretically demonstrated that snow lines can affect planetesimal formation in three ways. 1) Icy particles show increased sticking, favouring dust growth beyond the snow line location \citep[e.g.][]{Wada2009, Okuzumi2012}. However, CO has a lower dipole moment compared to more polar ices such as H$_2$O, which could actually lead to decreased sticking and growth beyond the CO snow line \citep[e.g.][]{Pinilla2017}. 2) Particles drifting inwards through the snow line lose their surface ice, causing a higher dust-to-gas ratio outside compared to just interior to the snow line \citep[e.g.][]{Stevenson1988,Cuzzi2004}. The evaporated gas may then diffuse beyond the snow line and freeze out onto incoming grains, leading to significantly enhanced growth at that location \citep{RosJohansen2013}, though further studies question the effectiveness of the latter process at the CO snow line \citep{Stammler2017}. 3) Sintering of icy particles can lead to enhanced fragmentation and, conversely, reduce growth just beyond an snow line \citep[e.g.][]{Okuzumi2016}. 

On the observational side, the emergence and abundance of concentric rings in recent observations of protoplanetary disks may indicate a variation in dust opacities at the snow line location of different species. However, this has been interpreted both ways as a sign of enhanced growth \citep{Zhang2015} or fragmentation \citep{Okuzumi2016}. 
Overall, it remains unclear whether the CO snow line would lead to an enhanced, reduced, or unchanged effectiveness of planetesimal formation. 

%The existence and abundance of planetesimal belts in the form of debris disks at tens of AU tells us that planetesimal formation does take place out to these distances. At the same time, planetesimals must have also formed (and further grown into planets) interior to these belts, as testified by both the abundance of exoplanets inwards of a few AU and by the architecture of our own Solar System. 
The tentative association between planetesimal belt locations and CO snow lines reported here could therefore indicate either of two scenarios. 1) Planetesimal formation is \textit{enhanced} at the CO snow line location, and is followed by rapid planet formation and inward migration. This mechanism could continue efficiently until the gas is dissipated, at which point the planetary system would be left with a belt of planetesimals that did not have time to further develop into planets just beyond the location of the CO snow line at the time of disk dispersal. A similar scenario has been proposed to explain the composition of Uranus and Neptune in the Solar System \citep{Ali-Dib2014}. 2) Planetesimal formation is \textit{inefficient} beyond the CO snow line location, leading to longer growth timescales which eventually allow planetesimals, but not planets, to form at these locations before the gas disk is dissipated. 

%The breadth of planetesimal belts has the potential to give further insights into planetesimal formation. Broad debris disks may be produced over time by the scattering action of a static or outward migrating planet interior to an originally narrow belt, but may also be born that way. The latter may be explained by the slow planetesimal formation timescales beyond the CO snow line, but potentially also by the ice line location changing over the protoplanetary disk lifetime \citep[e.g.][]{KennedyKenyon2008, Panic2017}. Conversely, a picture with all planetesimal belts born narrow at the CO snow line location would favour the enhanced growth scenario, unless fast radial drift can truncate the radial distribution of solids immediately beyond the CO snow line.
%\textbf{PARAGRAPH ABOUT VOLATILITY HERE - ALSO INCLUDE A SENTENCE IN NEXT PARAGRAPH AND MODIFY CONCLUSIONS SECTION.}
Regardless of whether the $R-L_{\star}$ relation for planetesimal belts is related to planetesimal formation at the CO snow line specifically, the similarity in slope between planetesimal belt and the two observed CO snow line locations would indicate that volatility of solids in protoplanetary disks plays a crucial role in planetesimal and/or planet formation. In turn, this could imply a broad similarity in cometary compositions across planetary systems, explaining ice abundances being so far consistent between exocomets and Solar System comets \citep{Matra2017a,Matra2017b,Matra2018a}.

\subsubsection{Inefficient planet formation}
\label{sect:planetform}

Another approach to understanding the origin of planetesimal belts is to consider why planetesimals did not go on to form planets, rather than why planetesimal themselves formed at specific locations in planetary systems. Given the known presence of planetary or brown dwarf mass companions interior to planetesimal belts \citep[e.g.][]{Marois2008,Lagrange2009,Rameau2013,Macintosh2015,Milli2017a}, and even a potential correlation between the two \citep{Wyatt2012,Kennedy2015}, a reasonable question to pose is whether planet formation simply did not have sufficient time to take place in the outer regions of planetary systems, where the mass budget is lower, and the orbital timescales are longer.
If that were to be the case, given that both the solid masses increase \citep[e.g.][]{Andrews2013} and the orbital periods shorten as function of stellar mass, it would make sense that planetesimal belts - which would be representative of the outer edge of planet formation - are observed to lie at larger radii around more massive (luminous) stars.
Using masses from Table \ref{tab:sourcelist}, our $R-L_{\star}$ relation translates in a similarly correlated $R-M_{\star}$ relation. Neglecting the effect of observational biases and collisional evolution, we find a power law dependence with slope $\alpha_{M_{\star}}\sim 1.0$ (i.e. $R\propto M_{\star}^{\alpha_{M_{\star}}}$).

Then, a simplified way to understand whether planet formation timescales could set this relation is to consider the accretion timescale for a protoplanet to reach a mass $M_{\rm pl}$ and radius $R_{\rm pl}$ through core accretion from a disk of planetesimals, and its dependence on $M_{\star}$. Following \citet{Kenyon2008}, this timescale can be estimated as $t\propto\frac{1}{\Sigma\Omega}$, where $\Sigma$ is the local surface density of planetesimals and $\Omega$ is the Keplerian angular frequency, where $\Omega\propto R^{-3/2}M_{\star}^{1/2}$. We assume a typical power-law planetesimal surface density profile ($\Sigma(R)\propto (M/R_{\mathrm{out}}^{y+2})R^y$) with total mass in planetesimals $M$ and extending from the star out to radius $R_{\mathrm{out}}$. We assume that the disk's average surface density is constant \citep[$R_{\mathrm{out}}^2\propto M$, as found for dust in protoplanetary disks,][]{Tripathi2017}, which implies ($\Sigma\propto M^{-0.5y}R^y$). We can then connect the total mass in planetesimals $M$ to the mass of the central star, assuming this dependence to be the same as observed for the dust mass in protoplanetary disks \citep[where $M_{\mathrm{dust}}\propto M_{\star}^x$, with $x\sim1.5-1.9$,][]{Pascucci2016}. 

Thus, if planetesimal accretion successfully produced planets out to a radius set by this accretion timescale, we would expect this radius to scale as $R\propto M_{\star}^{\frac{0.5(1-xy)}{1.5-y}}$.
If we assume a minimum mass solar nebula (MMSN) surface density profile with $y=-3/2$ \citep[][]{Weidenschilling1977, Hayashi1981}, the expectation would be that $R\propto M_{\star}^{0.54-0.64}$, which is shallower than the slope we reported here ($\alpha_{M_{\star}}\sim$1.0).%, and even more so compared to the slope we would obtain if we considered inner rather than average disk radii for planetesimal belts ($\alpha_{M_{\star}, \rm inner}\sim$1.2).

This simple calculation goes in the right direction to show that our result that planetesimal belt radii increase around stars of increasing mass qualitatively follows the expectation from a planet formation timescale perspective, although it produces a slope slightly shallower than observed. Furthermore, the timescales would not be quick enough, as this core accretion model cannot produce Uranus and Neptune in situ within the lifetime of the Solar Nebula \citep[e.g.][]{Goldreich2004}. 
A likely solution to several problems with this simple planetesimal accretion model has more recently been found through the pebble accretion model, where the growth rate is significantly sped up by the accretion of inward-drifting pebbles \citep[e.g.][]{LambrechtsJohansen2012,Bitsch2015}. Then, the timescale issue is overcome and several embryos can rapidly form in the outer regions of the Solar Nebula, and by extension in other planetary systems.
In terms of the $R-M_{\star}$ relation for planetesimal belts, pebble accretion would act to explain planet formation out to the inner edge of our observed relation in a shorter timescale. The pebble accretion rate and consequent planet formation timescale is highly dependent on the assumed protoplanetary disk parameters, making detailed comparison difficult. 

Given our main result that the scatter (rather than the slope or intercept) of resolved planetesimal belts is unlikely to be reproduced by current collisional evolution models and observational bias, perhaps a more important aspect to consider is how planet formation can reproduce such scatter. We speculate that this may be related to the range of timescales for planet formation in different planetary systems. If we let this timescale vary in the simple core accretion calculation above, we find $R\propto t^{1/(1.5-y)}$, where assuming $y=-3/2$ as for the MMSN yields $R\propto t^{1/3}$. The $\pm1\sigma$ scatter in radii found across the $R-L_{\star}$ relation (grey region in Fig. \ref{fig:mmlaw}) implies that $R^{+1\sigma}/R^{-1\sigma}\approx1.5$, which would imply a variation in planet formation timescales of $\sim3.4$. This would make sense if gas-rich protoplanetary disks producing detectable debris disks survived, for example, between $\sim$3-10 Myr, where these numbers are comparable to the observed decay in disk fraction in star-forming regions \citep[e.g.][]{Hernandez2008}.

Regardless of the details of the potential formation scenarios discussed here, confirming that there is a preferential formation location for planetesimal belts that is correlated with the host star's mass and luminosity would be important to provide one of the first extrasolar constraints to such planet formation models at large orbital separations. While confirmation requires expansion of the observational sample and a more complete model effort in the multi-dimensional parameter space of planetesimal belt observables, explaining its origin requires planet formation models and simulations to provide more specific predictions on the fate of planetesimals at large orbital separations, across a range of host star properties. At the same time, increasing the number of resolved snow lines in young protoplanetary disks, particularly across a range of stellar hosts, will also empirically contribute to confirming the potential link proposed here.

\section{Conclusions and Summary}
\label{sec:concl}
In this work, we collected radius measurements from all 26 extrasolar planetesimal belts resolved at millimetre wavelengths to date, and analysed their dependence on host star properties. We report the discovery of a statistically significant correlation between belt radii and host star luminosities, following $R=73^{+6}_{-6}L_{\star}^{0.19^{+0.04}_{-0.04}}$.

We simulate planetesimal belt populations to understand the effect of observational bias in $[R-L_{\star}]$ space. 
Given a static ring model, we show that it is unlikely that a population of belts with radii that are uncorrelated with the host star's luminosities can explain the observed $R-L_{\star}$ relation through selection effects alone. This is largely due to the observed population having a much lower scatter than the simulated one. 
We find the latter to remain true for several different sets of reasonable model assumptions, although we do not formally rule out that a specific combination of population model assumptions may explain the observed low scatter. Nonetheless, our tests indicates that an underlying $R-L_{\star}$ relation is likely necessary to explain the observed correlation. After repeating the fit to the observed population by taking into account observational bias through our fiducial model assumptions, we find the best-fit parameters of the relation to be largely unchanged, with $R=66.8^{+7.7}_{-11.8}L_{\star}^{0.19^{+0.05}_{-0.06}}$.

We then consider whether steady state collisional evolution of a population of belts that are once again uncorrelated in $[R-L_{\star}]$ space, coupled to observational bias, could explain the $R-L_{\star}$ relation. We do so by evolving the mass of simulated belt populations according to the models that fit the population of IR excesses \citep{Wyatt2007b, Sibthorpe2018}.
Including collisional evolution in the model population can readily explain the observed lack of small belts, particularly around stars of increasing luminosities. 
This brings the intrinsic scatter of the simulated population closer to the one observed, and better reproduces the observed $R-L_{\star}$ relation compared to a static population. However, the intrinsic scatter of the evolved simulated population is still higher and only marginally consistent with the one observed. This suggests that some of the collisional evolution model assumptions need to be refined; in particular, the $R-L_{\star}$ relation could indicate a preferential formation location for planetesimal belts in protoplanetary disks.

We briefly discuss how such a preferential formation location may be qualitatively explained in the context of current theories of planetesimal and planet formation. In particular, we focus on the CO snow line and its potential impact on the formation of planetesimals, showing that the location of the 2 observationally-determined CO snow lines in protoplanetary disks is close to the expectation from our $R-L_{\star}$ relation. The similar slope between planetesimal belts and CO snow lines would suggest that volatility is a driver of planetesimal and/or planet formation. 

At the same time, we consider why planetesimals did not grow further to form planets; we speculate that the inner edge of these belts may be set by the timescale of outermost planet formation, which would qualitatively explain the positive slope of the $R-L_{\star}$ relation. However, we find that this slope, in a simplified core accretion scenario, should be flatter than observed. The low scatter observed, on the other hand, may be due to a narrow range in planet formation timescales, and is in line with the expectation from core accretion and the range of observed protoplanetary disk lifetimes.

Our work shows that in order to shed more light on the origin of the $R-L_{\star}$ relation we need to expand the sample of resolved planetesimal belts, enabling simultaneous modelling of their masses and time evolution as well as radii distributions. This will be crucial in confirming that there is a preferential formation location of planetesimal belts, a finding that can set important new constraints on models of planetesimal and planet formation in the outer regions of the Solar System and extrasolar planetary systems.


%% Included in this acknowledgments section are examples of the
%% AASTeX hypertext markup commands. Use \url without the optional [HREF]
%% argument when you want to print the url directly in the text. Otherwise,
%% use either \url or \anchor, with the HREF as the first argument and the
%% text to be printed in the second.

\acknowledgments
LM acknowledges support from the Smithsonian Institution as a Submillimeter Array (SMA) Fellow. GMK is supported by the Royal Society as a Royal Society University
Research Fellow.
%We are grateful to V. Barger, T. Han, and R. J. N. Phillips for
%doing the math in section~\ref{bozomath}.
%More information on the AASTeX macros package is available \\ at
%\url{http://www.aas.org/publications/aastex}.
%For technical support, please write to
%\email{aastex-help@aas.org}.

%% facilities. A common set of keywords will make these types of searches>

%% copy editing.  Individual instruments can be provided in parentheses,
%% after the keyword, but they will not be verified.

\facility{SMA, ALMA}.

\bibliographystyle{apj}
\bibliography{lib}

\newcommand{\noop}[1]{}
\begin{thebibliography}{}
\expandafter\ifx\csname natexlab\endcsname\relax\def\natexlab#1{#1}\fi

\bibitem[{{Ali-Dib} {et~al.}(2014){Ali-Dib}, {Mousis}, {Petit}, \&
  {Lunine}}]{Ali-Dib2014}
{Ali-Dib}, M., {Mousis}, O., {Petit}, J.-M., \& {Lunine}, J.~I. 2014, \apj,
  793, 9

\bibitem[{{ALMA Partnership} {et~al.}(2015){ALMA Partnership}, {Brogan},
  {P{\'e}rez}, {Hunter}, {Dent}, {Hales}, {Hills}, {Corder}, {Fomalont},
  {Vlahakis}, {Asaki}, {Barkats}, {Hirota}, {Hodge}, {Impellizzeri}, {Kneissl},
  {Liuzzo}, {Lucas}, {Marcelino}, {Matsushita}, {Nakanishi}, {Phillips},
  {Richards}, {Toledo}, {Aladro}, {Broguiere}, {Cortes}, {Cortes}, {Espada},
  {Galarza}, {Garcia-Appadoo}, {Guzman-Ramirez}, {Humphreys}, {Jung}, {Kameno},
  {Laing}, {Leon}, {Marconi}, {Mignano}, {Nikolic}, {Nyman}, {Radiszcz},
  {Remijan}, {Rod{\'o}n}, {Sawada}, {Takahashi}, {Tilanus}, {Vila Vilaro},
  {Watson}, {Wiklind}, {Akiyama}, {Chapillon}, {de Gregorio-Monsalvo}, {Di
  Francesco}, {Gueth}, {Kawamura}, {Lee}, {Nguyen Luong}, {Mangum}, {Pietu},
  {Sanhueza}, {Saigo}, {Takakuwa}, {Ubach}, {van Kempen}, {Wootten},
  {Castro-Carrizo}, {Francke}, {Gallardo}, {Garcia}, {Gonzalez}, {Hill},
  {Kaminski}, {Kurono}, {Liu}, {Lopez}, {Morales}, {Plarre}, {Schieven},
  {Testi}, {Videla}, {Villard}, {Andreani}, {Hibbard}, \&
  {Tatematsu}}]{ALMAPartn2015}
{ALMA Partnership}, {Brogan}, C.~L., {P{\'e}rez}, L.~M., {et~al.} 2015, \apjl,
  808, L3

\bibitem[{{Andrews} {et~al.}(2013){Andrews}, {Rosenfeld}, {Kraus}, \&
  {Wilner}}]{Andrews2013}
{Andrews}, S.~M., {Rosenfeld}, K.~A., {Kraus}, A.~L., \& {Wilner}, D.~J. 2013,
  \apj, 771, 129

\bibitem[{{Andrews} {et~al.}(2016){Andrews}, {Wilner}, {Zhu}, {Birnstiel},
  {Carpenter}, {P{\'e}rez}, {Bai}, {{\"O}berg}, {Hughes}, {Isella}, \&
  {Ricci}}]{Andrews2016}
{Andrews}, S.~M., {Wilner}, D.~J., {Zhu}, Z., {et~al.} 2016, \apjl, 820, L40

\bibitem[{{Artymowicz} \& {Clampin}(1997)}]{ArtymowiczClampin1997}
{Artymowicz}, P., \& {Clampin}, M. 1997, \apj, 490, 863

\bibitem[{{Ballering} {et~al.}(2017){Ballering}, {Rieke}, {Su}, \&
  {G{\'a}sp{\'a}r}}]{Ballering2017}
{Ballering}, N.~P., {Rieke}, G.~H., {Su}, K.~Y.~L., \& {G{\'a}sp{\'a}r}, A.
  2017, \apj, 845, 120

\bibitem[{{Ballering} {et~al.}(2013){Ballering}, {Rieke}, {Su}, \&
  {Montiel}}]{Ballering2013}
{Ballering}, N.~P., {Rieke}, G.~H., {Su}, K.~Y.~L., \& {Montiel}, E. 2013,
  \apj, 775, 55

\bibitem[{{Baraffe} {et~al.}(2015){Baraffe}, {Homeier}, {Allard}, \&
  {Chabrier}}]{Baraffe2015}
{Baraffe}, I., {Homeier}, D., {Allard}, F., \& {Chabrier}, G. 2015, \aap, 577,
  A42

\bibitem[{{Birnstiel} {et~al.}(2016){Birnstiel}, {Fang}, \&
  {Johansen}}]{Birnstiel2016}
{Birnstiel}, T., {Fang}, M., \& {Johansen}, A. 2016, \ssr, 205, 41

\bibitem[{{Bitsch} {et~al.}(2015){Bitsch}, {Lambrechts}, \&
  {Johansen}}]{Bitsch2015}
{Bitsch}, B., {Lambrechts}, M., \& {Johansen}, A. 2015, \aap, 582, A112

\bibitem[{{Boccaletti} {et~al.}(2015){Boccaletti}, {Thalmann}, {Lagrange},
  {Janson}, {Augereau}, {Schneider}, {Milli}, {Grady}, {Debes}, {Langlois},
  {Mouillet}, {Henning}, {Dominik}, {Maire}, {Beuzit}, {Carson}, {Dohlen},
  {Engler}, {Feldt}, {Fusco}, {Ginski}, {Girard}, {Hines}, {Kasper}, {Mawet},
  {M{\'e}nard}, {Meyer}, {Moutou}, {Olofsson}, {Rodigas}, {Sauvage},
  {Schlieder}, {Schmid}, {Turatto}, {Udry}, {Vakili}, {Vigan}, {Wahhaj}, \&
  {Wisniewski}}]{Boccaletti2015}
{Boccaletti}, A., {Thalmann}, C., {Lagrange}, A.-M., {et~al.} 2015, \nat, 526,
  230

\bibitem[{{Booth} {et~al.}(2013){Booth}, {Kennedy}, {Sibthorpe}, {Matthews},
  {Wyatt}, {Duch{\^e}ne}, {Kavelaars}, {Rodriguez}, {Greaves}, {Koning},
  {Vican}, {Rieke}, {Su}, {Moro-Mart{\'{\i}}n}, \& {Kalas}}]{Booth2013}
{Booth}, M., {Kennedy}, G., {Sibthorpe}, B., {et~al.} 2013, \mnras, 428, 1263

\bibitem[{{Booth} {et~al.}(2016){Booth}, {Jord{\'a}n}, {Casassus}, {Hales},
  {Dent}, {Faramaz}, {Matr{\`a}}, {Barkats}, {Brahm}, \& {Cuadra}}]{Booth2016}
{Booth}, M., {Jord{\'a}n}, A., {Casassus}, S., {et~al.} 2016, \mnras, 460, L10

\bibitem[{{Booth} {et~al.}(2017){Booth}, {Dent}, {Jord{\'a}n}, {Lestrade},
  {Hales}, {Wyatt}, {Casassus}, {Ertel}, {Greaves}, {Kennedy}, {Matr{\`a}},
  {Augereau}, \& {Villard}}]{Booth2017}
{Booth}, M., {Dent}, W.~R.~F., {Jord{\'a}n}, A., {et~al.} 2017, \mnras, 469,
  3200

\bibitem[{{Bovy}(2017)}]{Bovy2017}
{Bovy}, J. 2017, \mnras, 470, 1360

\bibitem[{{Burns} {et~al.}(1979){Burns}, {Lamy}, \& {Soter}}]{Burns1979}
{Burns}, J.~A., {Lamy}, P.~L., \& {Soter}, S. 1979, \icarus, 40, 1

\bibitem[{{Casagrande} {et~al.}(2011){Casagrande}, {Sch{\"o}nrich}, {Asplund},
  {Cassisi}, {Ram{\'{\i}}rez}, {Mel{\'e}ndez}, {Bensby}, \&
  {Feltzing}}]{Casagrande2011}
{Casagrande}, L., {Sch{\"o}nrich}, R., {Asplund}, M., {et~al.} 2011, \aap, 530,
  A138

\bibitem[{{Casassus} {et~al.}(2013){Casassus}, {van der Plas}, {M}, {Dent},
  {Fomalont}, {Hagelberg}, {Hales}, {Jord{\'a}n}, {Mawet}, {M{\'e}nard},
  {Wootten}, {Wilner}, {Hughes}, {Schreiber}, {Girard}, {Ercolano}, {Canovas},
  {Rom{\'a}n}, \& {Salinas}}]{Casassus2013}
{Casassus}, S., {van der Plas}, G., {M}, S.~P., {et~al.} 2013, \nat, 493, 191

\bibitem[{{Chen} {et~al.}(2014){Chen}, {Mittal}, {Kuchner}, {Forrest}, {Lisse},
  {Manoj}, {Sargent}, \& {Watson}}]{Chen2014}
{Chen}, C.~H., {Mittal}, T., {Kuchner}, M., {et~al.} 2014, \apjs, 211, 25

\bibitem[{{Cuzzi} \& {Zahnle}(2004)}]{Cuzzi2004}
{Cuzzi}, J.~N., \& {Zahnle}, K.~J. 2004, \apj, 614, 490

\bibitem[{{Dent} {et~al.}(2014){Dent}, {Wyatt}, {Roberge}, {Augereau},
  {Casassus}, {Corder}, {Greaves}, {de Gregorio-Monsalvo}, {Hales}, {Jackson},
  {Hughes}, {Lagrange}, {Matthews}, \& {Wilner}}]{Dent2014}
{Dent}, W.~R.~F., {Wyatt}, M.~C., {Roberge}, A., {et~al.} 2014, Science, 343,
  1490

\bibitem[{{Eiroa} {et~al.}(2013){Eiroa}, {Marshall}, {Mora}, {Montesinos},
  {Absil}, {Augereau}, {Bayo}, {Bryden}, {Danchi}, {del Burgo}, {Ertel},
  {Fridlund}, {Heras}, {Krivov}, {Launhardt}, {Liseau}, {L{\"o}hne},
  {Maldonado}, {Pilbratt}, {Roberge}, {Rodmann}, {Sanz-Forcada}, {Solano},
  {Stapelfeldt}, {Th{\'e}bault}, {Wolf}, {Ardila}, {Ar{\'e}valo}, {Beichmann},
  {Faramaz}, {Gonz{\'a}lez-Garc{\'{\i}}a}, {Guti{\'e}rrez}, {Lebreton},
  {Mart{\'{\i}}nez-Arn{\'a}iz}, {Meeus}, {Montes}, {Olofsson}, {Su}, {White},
  {Barrado}, {Fukagawa}, {Gr{\"u}n}, {Kamp}, {Lorente}, {Morbidelli},
  {M{\"u}ller}, {Mutschke}, {Nakagawa}, {Ribas}, \& {Walker}}]{Eiroa2013}
{Eiroa}, C., {Marshall}, J.~P., {Mora}, A., {et~al.} 2013, \aap, 555, A11

\bibitem[{{Fedele} {et~al.}(2017){Fedele}, {Carney}, {Hogerheijde}, {Walsh},
  {Miotello}, {Klaassen}, {Bruderer}, {Henning}, \& {van
  Dishoeck}}]{Fedele2017}
{Fedele}, D., {Carney}, M., {Hogerheijde}, M.~R., {et~al.} 2017, \aap, 600, A72

\bibitem[{{Foreman-Mackey} {et~al.}(2013){Foreman-Mackey}, {Hogg}, {Lang}, \&
  {Goodman}}]{Foreman-Mackey2013}
{Foreman-Mackey}, D., {Hogg}, D.~W., {Lang}, D., \& {Goodman}, J. 2013, \pasp,
  125, 306

\bibitem[{{Goldreich} {et~al.}(2004){Goldreich}, {Lithwick}, \&
  {Sari}}]{Goldreich2004}
{Goldreich}, P., {Lithwick}, Y., \& {Sari}, R. 2004, \araa, 42, 549

\bibitem[{Goodman \& Weare(2010)}]{GoodmanWeare2010}
Goodman, J., \& Weare, J. 2010, Commun. Appl. Math. Comput. Sci., 5, 65

\bibitem[{{Greaves} {et~al.}(2014){Greaves}, {Sibthorpe}, {Acke}, {Pantin},
  {Vandenbussche}, {Olofsson}, {Dominik}, {Barlow}, {Bendo}, {Blommaert},
  {Brandeker}, {de Vries}, {Dent}, {Di Francesco}, {Fridlund}, {Gear},
  {Harvey}, {Hogerheijde}, {Holland}, {Ivison}, {Liseau}, {Matthews},
  {Pilbratt}, {Walker}, \& {Waelkens}}]{Greaves2014}
{Greaves}, J.~S., {Sibthorpe}, B., {Acke}, B., {et~al.} 2014, \apjl, 791, L11

\bibitem[{{Greaves} {et~al.}(2016){Greaves}, {Holland}, {Matthews}, {Marshall},
  {Dent}, {Woitke}, {Wyatt}, {Matr{\`a}}, \& {Jackson}}]{Greaves2016}
{Greaves}, J.~S., {Holland}, W.~S., {Matthews}, B.~C., {et~al.} 2016, \mnras,
  461, 3910

\bibitem[{{Hayashi}(1981)}]{Hayashi1981}
{Hayashi}, C. 1981, Progress of Theoretical Physics Supplement, 70, 35

\bibitem[{{Hern{\'a}ndez} {et~al.}(2005){Hern{\'a}ndez}, {Calvet}, {Hartmann},
  {Brice{\~n}o}, {Sicilia-Aguilar}, \& {Berlind}}]{Hernandez2005}
{Hern{\'a}ndez}, J., {Calvet}, N., {Hartmann}, L., {et~al.} 2005, \aj, 129, 856

\bibitem[{{Hern{\'a}ndez} {et~al.}(2008){Hern{\'a}ndez}, {Hartmann}, {Calvet},
  {Jeffries}, {Gutermuth}, {Muzerolle}, \& {Stauffer}}]{Hernandez2008}
{Hern{\'a}ndez}, J., {Hartmann}, L., {Calvet}, N., {et~al.} 2008, \apj, 686,
  1195

\bibitem[{{Holland} {et~al.}(2017){Holland}, {Matthews}, {Kennedy}, {Greaves},
  {Wyatt}, {Booth}, {Bastien}, {Bryden}, {Butner}, {Chen}, {Chrysostomou},
  {Davies}, {Dent}, {Di Francesco}, {Duchene}, {Gibb}, {Friberg}, {Ivison},
  {Jenness}, {Kavelaars}, {Lawler}, {Lestrade}, {Marshall}, {Moro-Martin},
  {Panic}, {Phillips}, {Serjeant}, {Schieven}, {Sibthorpe}, {Vican},
  {Ward-Thompson}, {van der Werf}, {White}, {Wilner}, \&
  {Zuckerman}}]{Holland2017}
{Holland}, W.~S., {Matthews}, B.~C., {Kennedy}, G.~M., {et~al.} 2017, ArXiv
  e-prints, arXiv:1706.01218

\bibitem[{{Hughes} {et~al.}(2017){Hughes}, {Lieman-Sifry}, {Flaherty}, {Daley},
  {Roberge}, {K{\'o}sp{\'a}l}, {Mo{\'o}r}, {Kamp}, {Wilner}, {Andrews},
  {Kastner}, \& {{\'A}brah{\'a}m}}]{Hughes2017}
{Hughes}, A.~M., {Lieman-Sifry}, J., {Flaherty}, K.~M., {et~al.} 2017, \apj,
  839, 86

\bibitem[{{Isella} {et~al.}(2016){Isella}, {Guidi}, {Testi}, {Liu}, {Li}, {Li},
  {Weaver}, {Boehler}, {Carperter}, {De Gregorio-Monsalvo}, {Manara}, {Natta},
  {P{\'e}rez}, {Ricci}, {Sargent}, {Tazzari}, \& {Turner}}]{Isella2016}
{Isella}, A., {Guidi}, G., {Testi}, L., {et~al.} 2016, Physical Review Letters,
  117, 251101

\bibitem[{{Jang-Condell} {et~al.}(2015){Jang-Condell}, {Chen}, {Mittal},
  {Manoj}, {Watson}, {Lisse}, {Nesvold}, \& {Kuchner}}]{Jang-Condell2015}
{Jang-Condell}, H., {Chen}, C.~H., {Mittal}, T., {et~al.} 2015, \apj, 808, 167

\bibitem[{{Janson} {et~al.}(2015){Janson}, {Quanz}, {Carson}, {Thalmann},
  {Lafreni{\`e}re}, \& {Amara}}]{Janson2015}
{Janson}, M., {Quanz}, S.~P., {Carson}, J.~C., {et~al.} 2015, \aap, 574, A120

\bibitem[{{Johansen} {et~al.}(2015){Johansen}, {Mac Low}, {Lacerda}, \&
  {Bizzarro}}]{Johansen2015}
{Johansen}, A., {Mac Low}, M.-M., {Lacerda}, P., \& {Bizzarro}, M. 2015,
  Science Advances, 1, 1500109

\bibitem[{{Jura}(1991)}]{Jura1991}
{Jura}, M. 1991, \apjl, 383, L79

\bibitem[{{Kains} {et~al.}(2011){Kains}, {Wyatt}, \& {Greaves}}]{Kains2011}
{Kains}, N., {Wyatt}, M.~C., \& {Greaves}, J.~S. 2011, \mnras, 414, 2486

\bibitem[{{Kelly}(2007)}]{Kelly2007}
{Kelly}, B.~C. 2007, \apj, 665, 1489

\bibitem[{{Kennedy} \& {Wyatt}(2010)}]{Kennedy2010}
{Kennedy}, G.~M., \& {Wyatt}, M.~C. 2010, \mnras, 405, 1253

\bibitem[{{Kennedy} \& {Wyatt}(2014)}]{KennedyWyatt2014}
---. 2014, \mnras, 444, 3164

\bibitem[{{Kennedy} {et~al.}(2015){Kennedy}, {Matr{\`a}}, {Marmier}, {Greaves},
  {Wyatt}, {Bryden}, {Holland}, {Lovis}, {Matthews}, {Pepe}, {Sibthorpe}, \&
  {Udry}}]{Kennedy2015}
{Kennedy}, G.~M., {Matr{\`a}}, L., {Marmier}, M., {et~al.} 2015, \mnras, 449,
  3121

\bibitem[{{Kenyon} \& {Bromley}(2008)}]{Kenyon2008}
{Kenyon}, S.~J., \& {Bromley}, B.~C. 2008, \apjs, 179, 451

\bibitem[{{Lagrange} {et~al.}(2009){Lagrange}, {Gratadour}, {Chauvin}, {Fusco},
  {Ehrenreich}, {Mouillet}, {Rousset}, {Rouan}, {Allard}, {Gendron}, {Charton},
  {Mugnier}, {Rabou}, {Montri}, \& {Lacombe}}]{Lagrange2009}
{Lagrange}, A.-M., {Gratadour}, D., {Chauvin}, G., {et~al.} 2009, \aap, 493,
  L21

\bibitem[{{Lambrechts} \& {Johansen}(2012)}]{LambrechtsJohansen2012}
{Lambrechts}, M., \& {Johansen}, A. 2012, \aap, 544, A32

\bibitem[{{Lewis}(1974)}]{Lewis1974}
{Lewis}, J.~S. 1974, Science, 186, 440

\bibitem[{{Lieman-Sifry} {et~al.}(2016){Lieman-Sifry}, {Hughes}, {Carpenter},
  {Gorti}, {Hales}, \& {Flaherty}}]{Lieman-Sifry2016}
{Lieman-Sifry}, J., {Hughes}, A.~M., {Carpenter}, J.~M., {et~al.} 2016, \apj,
  828, 25

\bibitem[{{Lissauer}(1987)}]{Lissauer1987}
{Lissauer}, J.~J. 1987, \icarus, 69, 249

\bibitem[{{Loomis} {et~al.}(2017){Loomis}, {{\"O}berg}, {Andrews}, \&
  {MacGregor}}]{Loomis2017}
{Loomis}, R.~A., {{\"O}berg}, K.~I., {Andrews}, S.~M., \& {MacGregor}, M.~A.
  2017, \apj, 840, 23

\bibitem[{{MacGregor} {et~al.}(2016{\natexlab{a}}){MacGregor}, {Lawler},
  {Wilner}, {Matthews}, {Kennedy}, {Booth}, \& {Di Francesco}}]{Macgregor2016b}
{MacGregor}, M.~A., {Lawler}, S.~M., {Wilner}, D.~J., {et~al.}
  2016{\natexlab{a}}, \apj, 828, 113

\bibitem[{{MacGregor} {et~al.}(2015){MacGregor}, {Wilner}, {Andrews}, \&
  {Hughes}}]{Macgregor2015a}
{MacGregor}, M.~A., {Wilner}, D.~J., {Andrews}, S.~M., \& {Hughes}, A.~M. 2015,
  \apj, 801, 59

\bibitem[{{MacGregor} {et~al.}(2013){MacGregor}, {Wilner}, {Rosenfeld},
  {Andrews}, {Matthews}, {Hughes}, {Booth}, {Chiang}, {Graham}, {Kalas},
  {Kennedy}, \& {Sibthorpe}}]{Macgregor2013}
{MacGregor}, M.~A., {Wilner}, D.~J., {Rosenfeld}, K.~A., {et~al.} 2013, \apjl,
  762, L21

\bibitem[{{MacGregor} {et~al.}(2016{\natexlab{b}}){MacGregor}, {Wilner},
  {Chandler}, {Ricci}, {Maddison}, {Cranmer}, {Andrews}, {Hughes}, \&
  {Steele}}]{Macgregor2016a}
{MacGregor}, M.~A., {Wilner}, D.~J., {Chandler}, C., {et~al.}
  2016{\natexlab{b}}, \apj, 823, 79

\bibitem[{{MacGregor} {et~al.}(2017){MacGregor}, {Matr{\`a}}, {Kalas},
  {Wilner}, {Pan}, {Kennedy}, {Wyatt}, {Duchene}, {Hughes}, {Rieke}, {Clampin},
  {Fitzgerald}, {Graham}, {Holland}, {Pani{\'c}}, {Shannon}, \&
  {Su}}]{Macgregor2017}
{MacGregor}, M.~A., {Matr{\`a}}, L., {Kalas}, P., {et~al.} 2017, \apj, 842, 8

\bibitem[{{Macintosh} {et~al.}(2015){Macintosh}, {Graham}, {Barman}, {De Rosa},
  {Konopacky}, {Marley}, {Marois}, {Nielsen}, {Pueyo}, {Rajan}, {Rameau},
  {Saumon}, {Wang}, {Patience}, {Ammons}, {Arriaga}, {Artigau}, {Beckwith},
  {Brewster}, {Bruzzone}, {Bulger}, {Burningham}, {Burrows}, {Chen}, {Chiang},
  {Chilcote}, {Dawson}, {Dong}, {Doyon}, {Draper}, {Duch{\^e}ne}, {Esposito},
  {Fabrycky}, {Fitzgerald}, {Follette}, {Fortney}, {Gerard}, {Goodsell},
  {Greenbaum}, {Hibon}, {Hinkley}, {Cotten}, {Hung}, {Ingraham},
  {Johnson-Groh}, {Kalas}, {Lafreniere}, {Larkin}, {Lee}, {Line}, {Long},
  {Maire}, {Marchis}, {Matthews}, {Max}, {Metchev}, {Millar-Blanchaer},
  {Mittal}, {Morley}, {Morzinski}, {Murray-Clay}, {Oppenheimer}, {Palmer},
  {Patel}, {Perrin}, {Poyneer}, {Rafikov}, {Rantakyr{\"o}}, {Rice}, {Rojo},
  {Rudy}, {Ruffio}, {Ruiz}, {Sadakuni}, {Saddlemyer}, {Salama}, {Savransky},
  {Schneider}, {Sivaramakrishnan}, {Song}, {Soummer}, {Thomas}, {Vasisht},
  {Wallace}, {Ward-Duong}, {Wiktorowicz}, {Wolff}, \&
  {Zuckerman}}]{Macintosh2015}
{Macintosh}, B., {Graham}, J.~R., {Barman}, T., {et~al.} 2015, Science, 350, 64

\bibitem[{Mamajek(2012)}]{Mamajek2012}
Mamajek, E.~E. 2012, The Astrophysical Journal, 754, L20

\bibitem[{{Mamajek} \& {Bell}(2014)}]{Mamajek2014}
{Mamajek}, E.~E., \& {Bell}, C.~P.~M. 2014, \mnras, 445, 2169

\bibitem[{{Mamajek} \& {Hillenbrand}(2008)}]{MamajekHillenbrand2008}
{Mamajek}, E.~E., \& {Hillenbrand}, L.~A. 2008, \apj, 687, 1264

\bibitem[{{Marino} {et~al.}(2015){Marino}, {Casassus}, {Perez}, {Lyra},
  {Roman}, {Avenhaus}, {Wright}, \& {Maddison}}]{Marino2015mwc758}
{Marino}, S., {Casassus}, S., {Perez}, S., {et~al.} 2015, \apj, 813, 76

\bibitem[{{Marino} {et~al.}(2017{\natexlab{a}}){Marino}, {Wyatt}, {Kennedy},
  {Holland}, {Matr{\`a}}, {Shannon}, \& {Ivison}}]{Marino2017b}
{Marino}, S., {Wyatt}, M.~C., {Kennedy}, G.~M., {et~al.} 2017{\natexlab{a}},
  \mnras, 469, 3518

\bibitem[{{Marino} {et~al.}(2016){Marino}, {Matr{\`a}}, {Stark}, {Wyatt},
  {Casassus}, {Kennedy}, {Rodriguez}, {Zuckerman}, {Perez}, {Dent}, {Kuchner},
  {Hughes}, {Schneider}, {Steele}, {Roberge}, {Donaldson}, \&
  {Nesvold}}]{Marino2016}
{Marino}, S., {Matr{\`a}}, L., {Stark}, C., {et~al.} 2016, \mnras, 460, 2933

\bibitem[{{Marino} {et~al.}(2017{\natexlab{b}}){Marino}, {Wyatt}, {Pani{\'c}},
  {Matr{\`a}}, {Kennedy}, {Bonsor}, {Kral}, {Dent}, {Duchene}, {Wilner},
  {Lisse}, {Lestrade}, \& {Matthews}}]{Marino2017a}
{Marino}, S., {Wyatt}, M.~C., {Pani{\'c}}, O., {et~al.} 2017{\natexlab{b}},
  \mnras, 465, 2595

\bibitem[{{Marois} {et~al.}(2008){Marois}, {Macintosh}, {Barman}, {Zuckerman},
  {Song}, {Patience}, {Lafreni{\`e}re}, \& {Doyon}}]{Marois2008}
{Marois}, C., {Macintosh}, B., {Barman}, T., {et~al.} 2008, Science, 322, 1348

\bibitem[{{Matr{\`a}} {et~al.}(2018){Matr{\`a}}, {Wilner}, {{\"O}berg},
  {Andrews}, {Loomis}, {Wyatt}, \& {Dent}}]{Matra2018a}
{Matr{\`a}}, L., {Wilner}, D.~J., {{\"O}berg}, K.~I., {et~al.} 2018, \apj, 853,
  147

\bibitem[{{Matr{\`a}} {et~al.}(2017{\natexlab{a}}){Matr{\`a}}, {Dent}, {Wyatt},
  {Kral}, {Wilner}, {Pani{\'c}}, {Hughes}, {de Gregorio-Monsalvo}, {Hales},
  {Augereau}, {Greaves}, \& {Roberge}}]{Matra2017a}
{Matr{\`a}}, L., {Dent}, W.~R.~F., {Wyatt}, M.~C., {et~al.} 2017{\natexlab{a}},
  \mnras, 464, 1415

\bibitem[{{Matr{\`a}} {et~al.}(2017{\natexlab{b}}){Matr{\`a}}, {MacGregor},
  {Kalas}, {Wyatt}, {Kennedy}, {Wilner}, {Duchene}, {Hughes}, {Pan}, {Shannon},
  {Clampin}, {Fitzgerald}, {Graham}, {Holland}, {Pani{\'c}}, \&
  {Su}}]{Matra2017b}
{Matr{\`a}}, L., {MacGregor}, M.~A., {Kalas}, P., {et~al.} 2017{\natexlab{b}},
  \apj, 842, 9

\bibitem[{{Matthews} {et~al.}(2014){Matthews}, {Krivov}, {Wyatt}, {Bryden}, \&
  {Eiroa}}]{Matthews2014}
{Matthews}, B.~C., {Krivov}, A.~V., {Wyatt}, M.~C., {Bryden}, G., \& {Eiroa},
  C. 2014, ArXiv e-prints, arXiv:1401.0743

\bibitem[{{Meshkat} {et~al.}(2013){Meshkat}, {Bailey}, {Rameau}, {Bonnefoy},
  {Boccaletti}, {Mamajek}, {Kenworthy}, {Chauvin}, {Lagrange}, {Su}, \&
  {Currie}}]{Meshkat2013}
{Meshkat}, T., {Bailey}, V., {Rameau}, J., {et~al.} 2013, \apjl, 775, L40

\bibitem[{{Milli} {et~al.}(2017){Milli}, {Hibon}, {Christiaens}, {Choquet},
  {Bonnefoy}, {Kennedy}, {Wyatt}, {Absil}, {G{\'o}mez Gonz{\'a}lez}, {del
  Burgo}, {Matr{\`a}}, {Augereau}, {Boccaletti}, {Delacroix}, {Ertel}, {Dent},
  {Forsberg}, {Fusco}, {Girard}, {Habraken}, {Huby}, {Karlsson}, {Lagrange},
  {Mawet}, {Mouillet}, {Perrin}, {Pinte}, {Pueyo}, {Reyes}, {Soummer},
  {Surdej}, {Tarricq}, \& {Wahhaj}}]{Milli2017a}
{Milli}, J., {Hibon}, P., {Christiaens}, V., {et~al.} 2017, \aap, 597, L2

\bibitem[{Mo\'{o}r {et~al.}(2013)Mo\'{o}r, Juh\'{a}sz, K\'{o}sp\'{a}l,
  \'{A}brah\'{a}m, Apai, Csengeri, Grady, Henning, Hughes, Kiss, Pascucci,
  Schmalzl, \& Gab\'{a}nyi}]{Moor2013}
Mo\'{o}r, A., Juh\'{a}sz, A., K\'{o}sp\'{a}l, A., {et~al.} 2013, 6

\bibitem[{{Mo{\'o}r} {et~al.}(2015){Mo{\'o}r}, {Henning}, {Juh{\'a}sz},
  {{\'A}brah{\'a}m}, {Balog}, {K{\'o}sp{\'a}l}, {Pascucci}, {Szab{\'o}},
  {Vavrek}, {Cur{\'e}}, {Csengeri}, {Grady}, {G{\"u}sten}, \&
  {Kiss}}]{Moor2015}
{Mo{\'o}r}, A., {Henning}, T., {Juh{\'a}sz}, A., {et~al.} 2015, \apj, 814, 42

\bibitem[{{Mo{\'o}r} {et~al.}(2017){Mo{\'o}r}, {Cur{\'e}}, {K{\'o}sp{\'a}l},
  {{\'A}brah{\'a}m}, {Csengeri}, {Eiroa}, {Gunawan}, {Henning}, {Hughes},
  {Juh{\'a}sz}, {Pawellek}, \& {Wyatt}}]{Moor2017}
{Mo{\'o}r}, A., {Cur{\'e}}, M., {K{\'o}sp{\'a}l}, {\'A}., {et~al.} 2017, ArXiv
  e-prints, arXiv:1709.08414

\bibitem[{{Morales} {et~al.}(2016){Morales}, {Bryden}, {Werner}, \&
  {Stapelfeldt}}]{Morales2016}
{Morales}, F.~Y., {Bryden}, G., {Werner}, M.~W., \& {Stapelfeldt}, K.~R. 2016,
  \apj, 831, 97

\bibitem[{{Natta} {et~al.}(2004){Natta}, {Testi}, {Neri}, {Shepherd}, \&
  {Wilner}}]{Natta2004}
{Natta}, A., {Testi}, L., {Neri}, R., {Shepherd}, D.~S., \& {Wilner}, D.~J.
  2004, \aap, 416, 179

\bibitem[{{Okuzumi} {et~al.}(2016){Okuzumi}, {Momose}, {Sirono}, {Kobayashi},
  \& {Tanaka}}]{Okuzumi2016}
{Okuzumi}, S., {Momose}, M., {Sirono}, S.-i., {Kobayashi}, H., \& {Tanaka}, H.
  2016, \apj, 821, 82

\bibitem[{{Okuzumi} {et~al.}(2012){Okuzumi}, {Tanaka}, {Kobayashi}, \&
  {Wada}}]{Okuzumi2012}
{Okuzumi}, S., {Tanaka}, H., {Kobayashi}, H., \& {Wada}, K. 2012, \apj, 752,
  106

\bibitem[{{Olofsson} {et~al.}(2016){Olofsson}, {Samland}, {Avenhaus},
  {Caceres}, {Henning}, {Mo{\'o}r}, {Milli}, {Canovas}, {Quanz}, {Schreiber},
  {Augereau}, {Bayo}, {Bazzon}, {Beuzit}, {Boccaletti}, {Buenzli}, {Casassus},
  {Chauvin}, {Dominik}, {Desidera}, {Feldt}, {Gratton}, {Janson}, {Lagrange},
  {Langlois}, {Lannier}, {Maire}, {Mesa}, {Pinte}, {Rouan}, {Salter},
  {Thalmann}, \& {Vigan}}]{Olofsson2016}
{Olofsson}, J., {Samland}, M., {Avenhaus}, H., {et~al.} 2016, \aap, 591, A108

\bibitem[{{Pascucci} {et~al.}(2016){Pascucci}, {Testi}, {Herczeg}, {Long},
  {Manara}, {Hendler}, {Mulders}, {Krijt}, {Ciesla}, {Henning}, {Mohanty},
  {Drabek-Maunder}, {Apai}, {Sz{\H u}cs}, {Sacco}, \&
  {Olofsson}}]{Pascucci2016}
{Pascucci}, I., {Testi}, L., {Herczeg}, G.~J., {et~al.} 2016, \apj, 831, 125

\bibitem[{{Pawellek} \& {Krivov}(2015)}]{Pawellek2015}
{Pawellek}, N., \& {Krivov}, A.~V. 2015, \mnras, 454, 3207

\bibitem[{{Pawellek} {et~al.}(2014){Pawellek}, {Krivov}, {Marshall},
  {Montesinos}, {{\'A}brah{\'a}m}, {Mo{\'o}r}, {Bryden}, \&
  {Eiroa}}]{Pawellek2014}
{Pawellek}, N., {Krivov}, A.~V., {Marshall}, J.~P., {et~al.} 2014, \apj, 792,
  65

\bibitem[{{Pecaut} \& {Mamajek}(2013)}]{PecautMamajek2013}
{Pecaut}, M.~J., \& {Mamajek}, E.~E. 2013, \apjs, 208, 9

\bibitem[{{Pecaut} \& {Mamajek}(2016)}]{PecautMamajek2016}
---. 2016, \mnras, 461, 794

\bibitem[{{Pecaut} {et~al.}(2012){Pecaut}, {Mamajek}, \& {Bubar}}]{Pecaut2012}
{Pecaut}, M.~J., {Mamajek}, E.~E., \& {Bubar}, E.~J. 2012, \apj, 746, 154

\bibitem[{{Phillips} {et~al.}(2010){Phillips}, {Greaves}, {Dent}, {Matthews},
  {Holland}, {Wyatt}, \& {Sibthorpe}}]{Phillips2010}
{Phillips}, N.~M., {Greaves}, J.~S., {Dent}, W.~R.~F., {et~al.} 2010, \mnras,
  403, 1089

\bibitem[{{Pinilla} {et~al.}(2017){Pinilla}, {Pohl}, {Stammler}, \&
  {Birnstiel}}]{Pinilla2017}
{Pinilla}, P., {Pohl}, A., {Stammler}, S.~M., \& {Birnstiel}, T. 2017, \apj,
  845, 68

\bibitem[{{Qi} {et~al.}(2015){Qi}, {{\"O}berg}, {Andrews}, {Wilner}, {Bergin},
  {Hughes}, {Hogherheijde}, \& {D'Alessio}}]{Qi2015}
{Qi}, C., {{\"O}berg}, K.~I., {Andrews}, S.~M., {et~al.} 2015, \apj, 813, 128

\bibitem[{{Qi} {et~al.}(2013){Qi}, {{\"O}berg}, {Wilner}, {D'Alessio},
  {Bergin}, {Andrews}, {Blake}, {Hogerheijde}, \& {van Dishoeck}}]{Qi2013}
{Qi}, C., {{\"O}berg}, K.~I., {Wilner}, D.~J., {et~al.} 2013, Science, 341, 630

\bibitem[{{Rameau} {et~al.}(2013){Rameau}, {Chauvin}, {Lagrange}, {Boccaletti},
  {Quanz}, {Bonnefoy}, {Girard}, {Delorme}, {Desidera}, {Klahr}, {Mordasini},
  {Dumas}, \& {Bonavita}}]{Rameau2013}
{Rameau}, J., {Chauvin}, G., {Lagrange}, A.-M., {et~al.} 2013, \apjl, 772, L15

\bibitem[{{Ricci} {et~al.}(2015){Ricci}, {Carpenter}, {Fu}, {Hughes}, {Corder},
  \& {Isella}}]{Ricci2015a}
{Ricci}, L., {Carpenter}, J.~M., {Fu}, B., {et~al.} 2015, \apj, 798, 124

\bibitem[{{Ros} \& {Johansen}(2013)}]{RosJohansen2013}
{Ros}, K., \& {Johansen}, A. 2013, \aap, 552, A137

\bibitem[{{Schwarz} {et~al.}(2016){Schwarz}, {Bergin}, {Cleeves}, {Blake},
  {Zhang}, {{\"O}berg}, {van Dishoeck}, \& {Qi}}]{Schwarz2016}
{Schwarz}, K.~R., {Bergin}, E.~A., {Cleeves}, L.~I., {et~al.} 2016, \apj, 823,
  91

\bibitem[{{Sibthorpe} {et~al.}(2018){Sibthorpe}, {Kennedy}, {Wyatt},
  {Lestrade}, {Greaves}, {Matthews}, \& {Duch{\^e}ne}}]{Sibthorpe2018}
{Sibthorpe}, B., {Kennedy}, G.~M., {Wyatt}, M.~C., {et~al.} 2018, \mnras, 475,
  3046

\bibitem[{{Sierchio} {et~al.}(2014){Sierchio}, {Rieke}, {Su}, \&
  {G{\'a}sp{\'a}r}}]{Sierchio2014}
{Sierchio}, J.~M., {Rieke}, G.~H., {Su}, K.~Y.~L., \& {G{\'a}sp{\'a}r}, A.
  2014, \apj, 785, 33

\bibitem[{{Spangler} {et~al.}(2001){Spangler}, {Sargent}, {Silverstone},
  {Becklin}, \& {Zuckerman}}]{Spangler2001}
{Spangler}, C., {Sargent}, A.~I., {Silverstone}, M.~D., {Becklin}, E.~E., \&
  {Zuckerman}, B. 2001, \apj, 555, 932

\bibitem[{{Stammler} {et~al.}(2017){Stammler}, {Birnstiel}, {Pani{\'c}},
  {Dullemond}, \& {Dominik}}]{Stammler2017}
{Stammler}, S.~M., {Birnstiel}, T., {Pani{\'c}}, O., {Dullemond}, C.~P., \&
  {Dominik}, C. 2017, \aap, 600, A140

\bibitem[{{Steele} {et~al.}(2016){Steele}, {Hughes}, {Carpenter}, {Ricarte},
  {Andrews}, {Wilner}, \& {Chiang}}]{Steele2016}
{Steele}, A., {Hughes}, A.~M., {Carpenter}, J., {et~al.} 2016, \apj, 816, 27

\bibitem[{{Stern} \& {Colwell}(1997)}]{Stern1997}
{Stern}, S.~A., \& {Colwell}, J.~E. 1997, \apj, 490, 879

\bibitem[{{Stevenson} \& {Lunine}(1988)}]{Stevenson1988}
{Stevenson}, D.~J., \& {Lunine}, J.~I. 1988, \icarus, 75, 146

\bibitem[{{Strubbe} \& {Chiang}(2006)}]{StrubbeChiang2006}
{Strubbe}, L.~E., \& {Chiang}, E.~I. 2006, \apj, 648, 652

\bibitem[{{Su} {et~al.}(2006){Su}, {Rieke}, {Stansberry}, {Bryden},
  {Stapelfeldt}, {Trilling}, {Muzerolle}, {Beichman}, {Moro-Martin}, {Hines},
  \& {Werner}}]{Su2006}
{Su}, K.~Y.~L., {Rieke}, G.~H., {Stansberry}, J.~A., {et~al.} 2006, \apj, 653,
  675

\bibitem[{{Su} {et~al.}(2017){Su}, {Macgregor}, {Booth}, {Wilner}, {Flaherty},
  {Hughes}, {Phillips}, {Malhotra}, {Hales}, {Morrison}, {Ertel}, {Matthews},
  {Dent}, \& {Casassus}}]{Su2017}
{Su}, K.~Y.~L., {Macgregor}, M.~A., {Booth}, M., {et~al.} 2017, ArXiv e-prints,
  arXiv:1709.10129

\bibitem[{{Tielens}(2005)}]{Tielens2005}
{Tielens}, A.~G.~G.~M. 2005, {The Physics and Chemistry of the Interstellar
  Medium}

\bibitem[{{Tripathi} {et~al.}(2017){Tripathi}, {Andrews}, {Birnstiel}, \&
  {Wilner}}]{Tripathi2017}
{Tripathi}, A., {Andrews}, S.~M., {Birnstiel}, T., \& {Wilner}, D.~J. 2017,
  \apj, 845, 44

\bibitem[{{Valenti} \& {Fischer}(2005)}]{Valenti2005}
{Valenti}, J.~A., \& {Fischer}, D.~A. 2005, \apjs, 159, 141

\bibitem[{{van der Marel} {et~al.}(2013){van der Marel}, {van Dishoeck},
  {Bruderer}, {Birnstiel}, {Pinilla}, {Dullemond}, {van Kempen}, {Schmalzl},
  {Brown}, {Herczeg}, {Mathews}, \& {Geers}}]{vanderMarel2013}
{van der Marel}, N., {van Dishoeck}, E.~F., {Bruderer}, S., {et~al.} 2013,
  Science, 340, 1199

\bibitem[{{Wada} {et~al.}(2009){Wada}, {Tanaka}, {Suyama}, {Kimura}, \&
  {Yamamoto}}]{Wada2009}
{Wada}, K., {Tanaka}, H., {Suyama}, T., {Kimura}, H., \& {Yamamoto}, T. 2009,
  \apj, 702, 1490

\bibitem[{{Webb} {et~al.}(1999){Webb}, {Zuckerman}, {Platais}, {Patience},
  {White}, {Schwartz}, \& {McCarthy}}]{Webb1999}
{Webb}, R.~A., {Zuckerman}, B., {Platais}, I., {et~al.} 1999, \apjl, 512, L63

\bibitem[{{Weidenschilling}(1977)}]{Weidenschilling1977}
{Weidenschilling}, S.~J. 1977, \apss, 51, 153

\bibitem[{{Wyatt}(2006)}]{Wyatt2006}
{Wyatt}, M.~C. 2006, \apj, 639, 1153

\bibitem[{Wyatt(2008)}]{Wyatt2008}
Wyatt, M.~C. 2008, Annual Review of Astronomy and Astrophysics, 46, 339

\bibitem[{{Wyatt} {et~al.}(2015){Wyatt}, {Pani{\'c}}, {Kennedy}, \&
  {Matr{\`a}}}]{Wyatt2015}
{Wyatt}, M.~C., {Pani{\'c}}, O., {Kennedy}, G.~M., \& {Matr{\`a}}, L. 2015,
  \apss, 357, 103

\bibitem[{{Wyatt} {et~al.}(2007){Wyatt}, {Smith}, {Su}, {Rieke}, {Greaves},
  {Beichman}, \& {Bryden}}]{Wyatt2007b}
{Wyatt}, M.~C., {Smith}, R., {Su}, K.~Y.~L., {et~al.} 2007, \apj, 663, 365

\bibitem[{Wyatt {et~al.}(2012)Wyatt, Kennedy, Sibthorpe, Moro-Mart\'{\i}n,
  Lestrade, Ivison, Matthews, Udry, Greaves, Kalas, Lawler, Su, Rieke, Booth,
  Bryden, Horner, Kavelaars, \& Wilner}]{Wyatt2012}
Wyatt, M.~C., Kennedy, G., Sibthorpe, B., {et~al.} 2012, Monthly Notices of the
  Royal Astronomical Society, 424, 1206

\bibitem[{{Zhang} {et~al.}(2015){Zhang}, {Blake}, \& {Bergin}}]{Zhang2015}
{Zhang}, K., {Blake}, G.~A., \& {Bergin}, E.~A. 2015, \apjl, 806, L7

\bibitem[{{Zuckerman} {et~al.}(2011){Zuckerman}, {Rhee}, {Song}, \&
  {Bessell}}]{Zuckerman2011}
{Zuckerman}, B., {Rhee}, J.~H., {Song}, I., \& {Bessell}, M.~S. 2011, \apj,
  732, 61

\end{thebibliography}

\appendix

\section{Analytical estimates of the optical depth for face-on and edge-on narrow belts}
\label{sec:optdepder}

The definition of optical depth for a column of dust of length $L$ (neglecting scattering) at a wavelength $\lambda$ reads \citep[e.g.][]{Tielens2005}
\begin{equation}
\tau(\lambda)=L\int_{a_{\rm min}}^{a_{\mathrm{max}}}n_{\mathrm{d}}(a)\sigma_{\rm d}(a)Q_{\rm abs}(a, \lambda)da,
\end{equation}
where $a_{\mathrm{max}}$ and $a_{\mathrm{min}}$ are the minimum and maximum sizes of the dust distribution, $n_{\mathrm{d}}$ is the number density of dust grains, $\sigma_{\mathrm{d}}$ is the cross-sectional area of a single dust grain, and $Q_{\mathrm{abs}}$ is the grain's absorption efficiency, which is size and wavelength dependent. We assume that continuum emission at any wavelength is dominated by grains of the same size as the wavelength of the emission, leading to the approximation $Q_{\rm abs}\sim 1$.
Then, given that $\sigma_{\rm tot}\equiv\int_{a_{\rm min}}^{a_{\mathrm{max}}}N_{\mathrm{d}}(a)da$ (where $N$ is the number, rather than number density, of dust grains of size $a$) the optical depth through the column integrated over all emission wavelengths can be found as $\tau=\frac{\sigma_{\rm tot}L}{V}$,
where $V$ is the volume of the dust column with line of sight length $L$.

For a narrow belt approximated as a box of height $H$ and width $\Delta R$ (with uniform dust number density) observed face-on, the on-sky area $V/L$ can be estimated as $2\pi R\Delta R$, which combined to the definition of fractional luminosity $f=\sigma_{\rm tot}/(4\pi R^2)$ leads to
\begin{equation}
\tau_{\rm face-on}=\frac{2f}{\Delta R/R}.
\end{equation}
This implies that a belt with $\Delta R/R$ of 0.5 becomes optically thick ($\tau_{\rm face-on}>1$) if its fractional luminosity is greater than $2.5\times10^{-1}$ (as argued in \S\ref{sec:optthick}).

For the same narrow belt in the uniform density box model approximation, we also consider its optical depth in the perfectly edge-on viewing scenario. In this case, the maximum optical depth is attained along the column of maximum length $L_{\rm max}$ along the line of sight through the disk. This column corresponds to the tangent to the inner radius of the belt along the line of sight, leading to $L_{\rm max}=2\sqrt{2R\Delta R}=2R\sqrt{2\Delta R/R}$. For a uniform density box-like ring, the volume corresponds to $V=2\pi R^2(\Delta R/R)H$. Then, the maximum optical depth for the edge-on belt can be estimated as
\begin{equation}
\tau_{\rm edge-on, max}=\frac{4\pi Rf}{H}\sqrt{\frac{2R}{\pi\Delta R}},
\end{equation}
leading to the conclusion in \S\ref{sec:optthick} that a belt with aspect ratio $H/R=0.1$ and $\Delta R/R=0.5$ only becomes optically thick for fractional luminosities $f>7.1\times10^{-3}$.

\section{The number of resolved belts in $[R,L_{\star}]$ space}
\label{sec:resbeltab}

\subsection{Static model}

\begin{figure*}
\vspace{-0mm}
 \hspace{-0mm}
  \includegraphics*[scale=0.49]{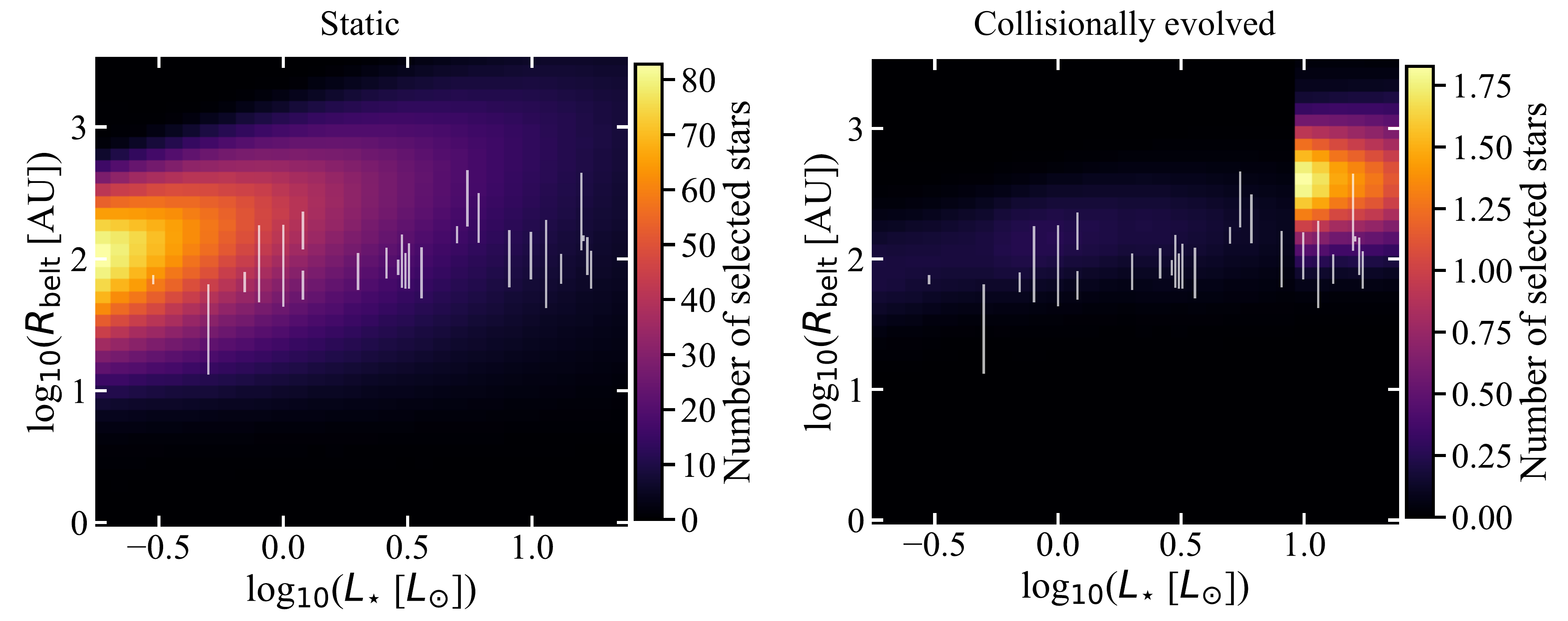}
\vspace{-2mm}
\caption{\textit{Left:} Number of stars expected to have detectable and mm-resolvable belts as a function of radius and stellar luminosity within 150pc from Earth, per $\rm log_{10}$$([R,L_{\star}])$ bin. We assumed that all stars have been observed out to 150 pc, and that they all host a belt that does not evolve with time. The same fiducial model population as Fig. \ref{fig:detectmap} has been assumed; the difference is that this figure takes into account the stellar density and luminosity function of the Solar neighbourhood.
\textit{Right:} Result obtained after collisionally evolving the population in the left panel according to the model and best-fit parameters of \citet{Wyatt2007b} and \citet{Sibthorpe2018}. For a star of a given luminosity, its belt is evolved to an age that is drawn from a uniform distribution up to the smallest between its main-sequence lifetime and the age of the universe. Compared to Fig. \ref{fig:detectmap_evolved}, this considers the stellar density and luminosity function of the Solar neighbourhood, but also the distribution of belt radii ($N(R)\propto R^{\gamma}$) assumed by the evolutionary model.
%\textit{Left:} \textbf{FIX THIS} Selection probability (\%) per $\rm log_{10}$$([R,L_{\star}])$ bin of a simulated population of belts whose dust mass has been evolved according to the model and best-fit parameters of \citet{Wyatt2007b} and \citet{Sibthorpe2018}. \textbf{For a star of a given luminosity, an age is drawn from a uniform distribution up to the smallest between its main-sequence lifetime and the age of the universe.} Belt distances from Earth are randomly drawn from an isotropic distribution ($N(d)\propto d^2$) out to 150 pc. White vertical bars represent our sample of belts currently resolved at millimeter wavelengths from Fig. \ref{fig:mmlaw}. \textit{Right:} Number of stars expected to have detectable and mm-resolvable belts according to this collisionally evolved model population, as a function of radius and stellar luminosity within 150pc from Earth, per $\rm log_{10}$$([R,L_{\star}])$ bin. All stars have been assumed to host a belt. Compared to the left panel, the right panel takes into account the stellar density and luminosity function of the Solar neighbourhood, as well as the distribution of belt radii ($N(R)\propto R^{\gamma}$) assumed by the evolutionary model.} 
}
\label{fig:nstars}
\end{figure*}

As mentioned in Sect. \ref{sec:undersbias}, our selection probability map in Fig. \ref{fig:detectmap} does not consider the stellar luminosity function in the Solar neighbourhood; in other words, the detection fraction for each stellar luminosity does not take into account that lower luminosity stars are much more abundant than higher luminosity stars. The latter is needed to be able to consider the number of stars - as opposed to their detection fraction - expected to have a detectable and resolvable belt in any $[R, L_{\star}]$ bin. In order to do this, we turn the selection fraction per $[R, L_{\star}]$ bin (Fig. \ref{fig:detectmap}) into a selection fraction per $L_{\star}$ column, and multiply the latter by the number of stars of that luminosity within 150 pc from Earth. This number of stars is calculated using local stellar densities as a function of spectral type from \citet{Bovy2017}, and assuming uniform stellar density. Since these stellar density measurements only extend down to K4 spectral type, in this step we only consider stars more luminous than K4 ($L_{\star}\geq0.18$ L$_{\odot}$).

The resulting map is shown in Fig. \ref{fig:nstars} (left), and shows significant differences compared to the selection fraction map in Fig. \ref{fig:detectmap}. While the selection fraction is higher around more luminous stars, the predicted absolute number of stars selected is higher for less luminous stars. This is readily attributable to the stellar luminosity function favouring less luminous stars. At the same time, we find that the lower envelope of detectability now increases as a function of stellar luminosity. This is because for a given belt radius, there will be a lot more low luminosity stars at a distance $d$ close to Earth, which in turn implies more low-luminosity belts with higher fluxes that are more easily detectable. Then, regardless of the belt radius, our fiducial model applied to the stellar population in the Solar neighbourhood predicts that the observed abundance of selected belts should favour lower luminosity stars.

Instead, the number of stars in our resolved sample is found to increase with stellar luminosity. This could be due in part to some selection bias not taken into account here (for example, the survey of \citet{Moor2017} was specifically targeted at A-type stars), but also to the fact that we so-far assumed the set of parameters $[f, R/R_{\rm BB}, \lambda_0, \beta]$ to be independent of stellar luminosity. A decreasing $R/R_{\rm BB}$ with $L_{\star}$ as reported from the \textit{Herschel} resolved disks \citep{Pawellek2014} would increase the detectability of disks around lower luminosity stars even further, as disks around less luminous stars, for the same radius, would be hotter and brighter. Additionally, we see no significant correlation between $R/R_{\rm BB}$ and $L_{\star}$ in our observed sample; therefore, an $R/R_{\rm BB}$ dependence on $L_{\star}$ cannot explain this trend.

On the other hand, we deem it is plausible that the observed increase in number of resolved belts around more luminous stars is caused by these belts having higher fractional luminosities, a trend that is tentatively present in the observed sample. This could be explained by the fact that more luminous stars are also, on average, younger, implying that they had less time to collisionally deplete. This could make them brighter, potentially explaining the increasing abundance of resolved belts around more luminous stars; we explore this possibility when considering collisional evolution in \ref{sect:nstarscollev}.

%The aim of this work, however, is to investigate the origin of the observed $R(L_{\star})$ relation rather than the population density in $[R, L_{\star}]$ space, $N_{\rm belts}(R, L_{\star})$. 

\subsection{Collisionally evolved model}
\label{sect:nstarscollev}

The right panel of Fig. \ref{fig:nstars} shows an equivalent map of $N(R, L_{\star})$ of selected belts after the model population has been collisionally evolved according to the model of \citet{Wyatt2007b} and \citet{Sibthorpe2018}. In this case, while including the stellar luminosity function $N(L_{\star})$ favours low luminosity stars (as found for the static population in the left panel of the figure), including the distribution of belt blackbody radii $N(R_{\rm BB})$ assumed by this model favours A stars, as the A star population was best fit by a power law with a much flatter slope compared to FGK stars ($\gamma=-0.8$ versus $\gamma=-1.7$). For A stars, this means favouring larger disks, which evolve more slowly and are thus more easily detectable. %Overall, the predicted number of disks as a function of stellar luminosity shows a dip for F and early G stars, which is not observed. This is likely caused by the same problem seen in the static model, i.e. that the model predicts too many belts around low luminosity stars. As mentioned in \S\ref{sec:resbeltab}, that could be due to an age bias between stars of different luminosities, which is not taken into account here (as we have evolved all stars to the same age of 100 Myr).

Overall, it appears that the collisional evolution model is able to produce belts at the radii where they are mostly observed, but fails to reproduce the number of stars having a detectable and resolvable belt as a function of stellar luminosity. This could be due to limitations of the model but also to some of our assumptions. For example, we are assuming that every star out to 150 pc has a disk, and that each disk has been targeted by IR observations. This is of course not the case, particularly for lower luminosity stars; for example, \textit{Herschel} surveys such as DEBRIS \citep{Phillips2010} and DUNES \citep{Eiroa2013} were designed to survey the same number of the nearest A, F, G, K, and M stars. As lower luminosity stars are more abundant, however, a much smaller fraction has been surveyed; this causes us to infer that there is a lower absolute number of disks around low luminosity stars than there truly is. Additionally, stars belonging to young moving groups or associations were likely preferentially targeted, introducing a bias toward younger stars.

We conclude that, as demonstrated above, the observed population density $N_{\rm belts}(R, L_{\star})$ has the potential to set useful constraints on some of our model parameters assumed, but is also exposed to other biases that are difficult to account for. In the most complete approach, all parameters determining a belt's observables including $R$ and $L_{\star}$ should be fitted simultaneously to $N_{\rm belts}(R, L_{\star}, f, R/R_{\rm BB}, \lambda_0, \beta, \rm age)$ in a comprehensive population study, which is however beyond the scope of this work. Since here we are not aiming to fit the population density but only the $R(L_{\star})$ relation, in Sect. \ref{sec:biasalone} and \ref{sec:steadystateevol} we opted to keep the stellar properties fixed to those of our observed population of 26 stars, and evaluate the likelihood of drawing our $[R, L_{\star}]$ relation from an underlying uncorrelated population \textit{given our sample of observed stars}.

\section{Testing different model assumptions for the bias analysis}
\label{sec:testass}

\begin{figure*}
%\begin{subfigure}{0.47\textwidth}
\vspace{-2mm}
% \centering
  \hspace{-11mm}
  \includegraphics*[scale=0.72]{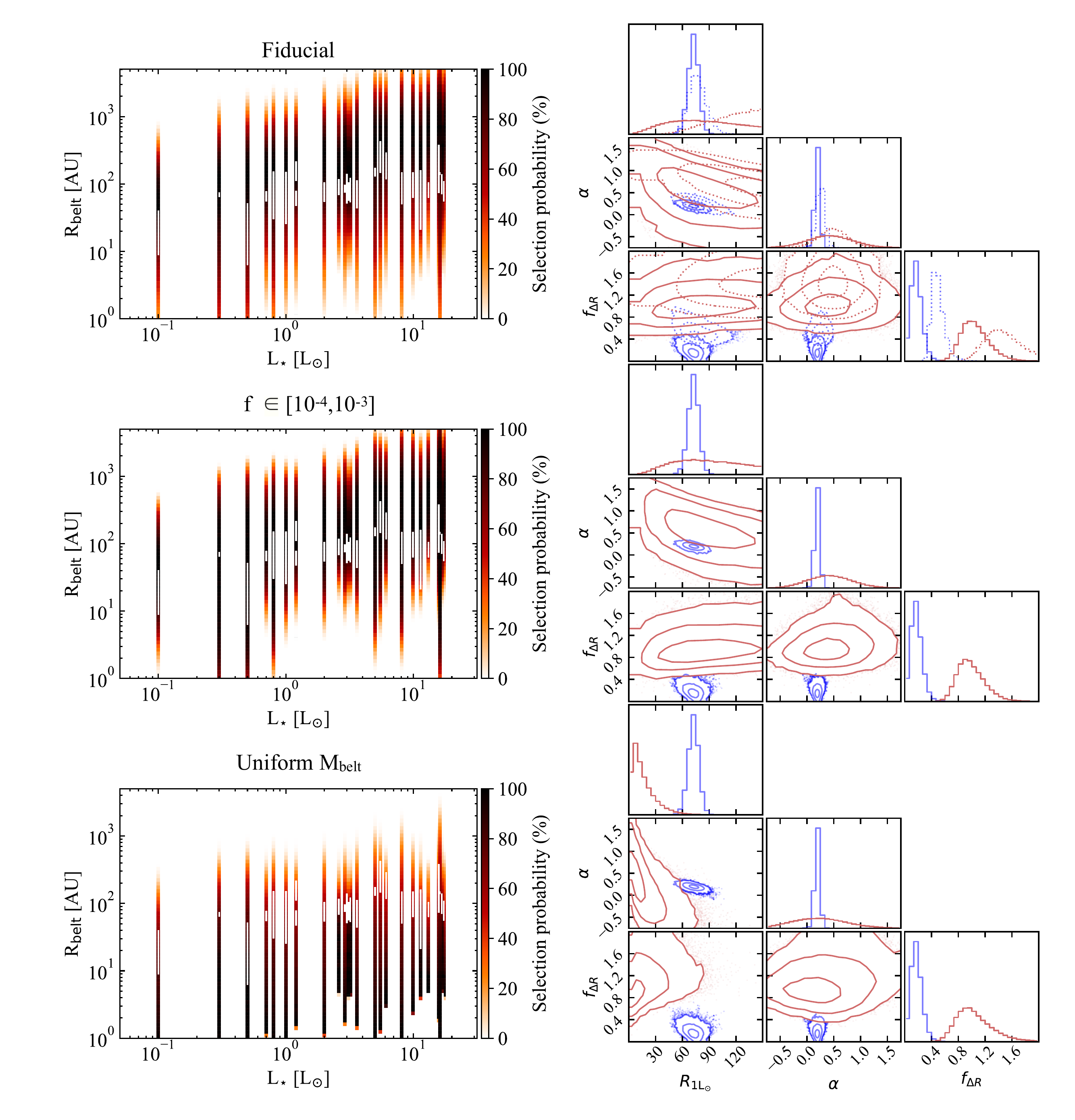}
%\vspace{0mm}
%\end{subfigure} \\
\vspace{-10mm}
\caption{Results of Monte Carlo simulations of model belt populations, where different rows show different model assumptions (see text for details). \textit{Left column:} Vertical color strips represent the normalized selection probability of belts in $\rm log_{10}$$(R)$ space for each of the stars in our sample, given their luminosity $L_{\star}$ and distance from Earth $d$. White vertical bars represent our sample of belts currently resolved at millimeter wavelengths from Fig. \ref{fig:mmlaw}. \textit{Right column:} Red solid histograms and contours represent marginalised probabilities of randomly drawing a given power law slope (left sub-column), intercept (centre sub-column) and intrinsic scatter (right sub-column) from an uncorrelated population of model belts, after accounting for observational selection effects. 1D histograms represent probability distributions of each parameter marginalised over the other two, whereas contour maps represent 2D probability distributions of different pairs of parameters, marginalised over the third. Contours represent the central [68.3, 95.5, 99.73] \% of the distribution. Blue solid lines represent marginalised posterior probability distributions of the parameters given the data, and should be compared with the model. Dotted lines represent probability distributions obtained when fixing the radius uncertainty to $\Delta R/R=0.1$ for all stars, and for both the observed and simulated data.} 
\label{fig:corner_uncorr_testass_1}
\end{figure*}

\begin{figure*}
%\begin{subfigure}{0.47\textwidth}
\vspace{-2mm}
% \centering
  \hspace{-11mm}
  \includegraphics*[scale=0.72]{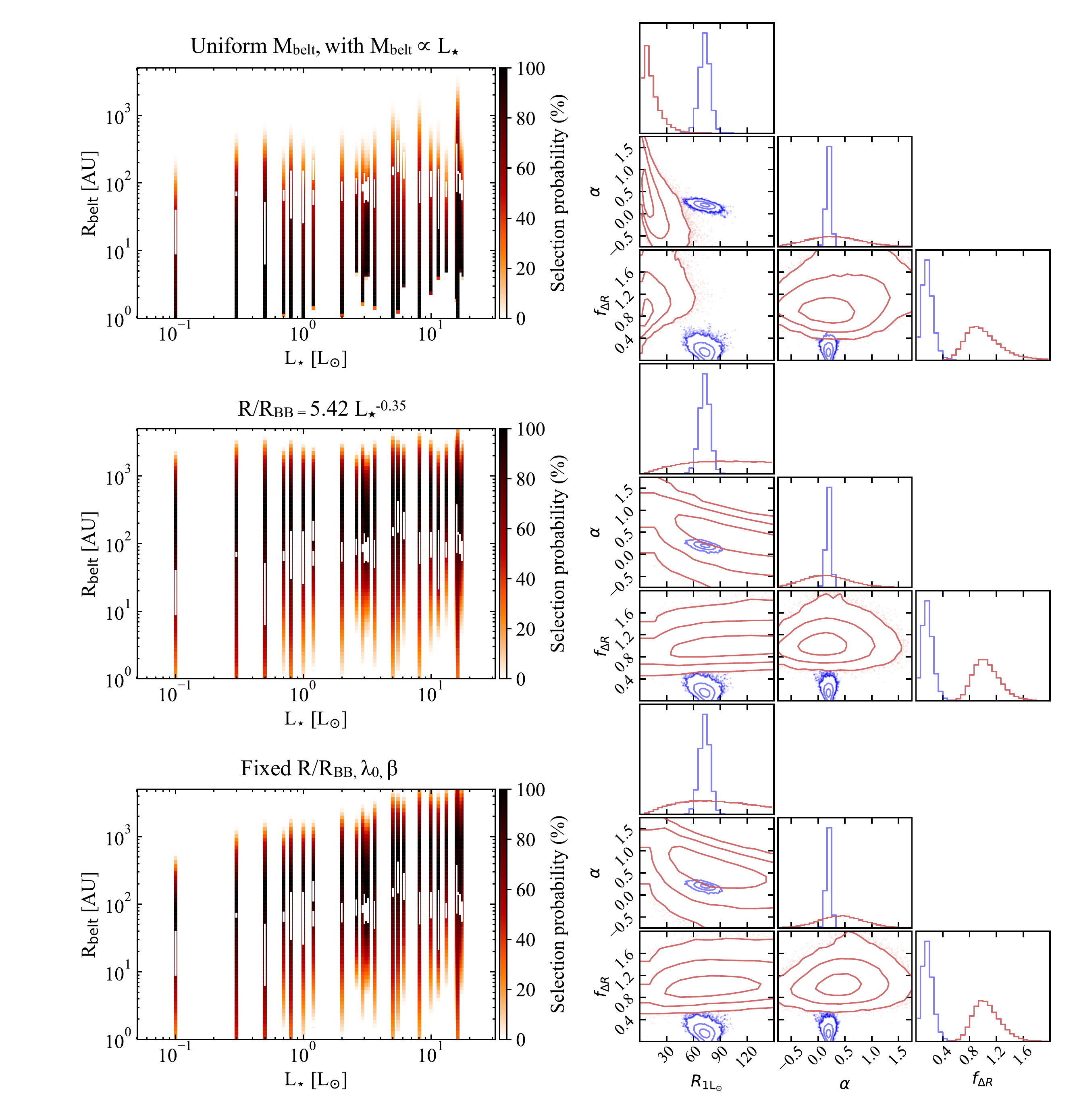}
%\vspace{0mm}
%\end{subfigure} \\
\vspace{-10mm}
\caption{Results of Monte Carlo simulations of model belt populations, where different rows show different model assumptions (see text for details). For both columns, lines and symbols have the same meaning as in Fig. \ref{fig:corner_uncorr_testass_1}.} 
\label{fig:corner_uncorr_testass_2}
\end{figure*}

The main limitation of the analysis in Sect. \ref{sec:biasalone} lies within our assumptions for the distributions of parameters [$f$, $R/R_{\rm BB}$, $\lambda_0$, $\beta$] for the simulated belt populations. This is because changing these parameters modifies the selection probability distribution of radii for each observed $L_{\star}$. We therefore repeat the process by changing the fiducial assumptions for the parameter distributions one at a time while keeping the others fixed. Figs. \ref{fig:corner_uncorr_testass_1} and \ref{fig:corner_uncorr_testass_2} (left columns) show how the radius selection probability distributions for each of our observed stellar luminosity vary after changing our assumptions from our fiducial model.

\textit{Changing the boundaries of the fractional luminosity distribution}. Maintaining a log-uniform distribution of fractional luminosities, we test the effect of changing the maximum and minimum boundaries of the simulated population. In particular, in Fig. \ref{fig:corner_uncorr_testass_1} (central row) we increased the lower boundary from 10$^{-7}$ to 10$^{-4}$, and lowered the upper boundary from 10$^{-2}$ to 10$^{-3}$. We find that the main effect of lowering the upper boundary is to push the lower limit of detectability to larger radii, making the vertical selection probability distribution narrower. On the other hand, increasing the lower boundary of the fractional luminosity distribution causes an overall increase in the selection probability for each star, but without changing the lower or upper limit of radius detectability. In practice, this means a wider range of radii `saturate' to a 100\% normalized selection probability (black in the colour scale of Fig. \ref{fig:corner_uncorr_testass_1}, left column).

When looking at the formal probability distributions, we find that the probability of drawing an intrinsic scatter $f_{\Delta R}$ within $\pm$1$\sigma$ of our observed value from an underlying uncorrelated population is marginally higher ($\sim2\times 10^{-5}$), but still very low. Changing the fractional luminosity boundaries to different values does not significantly improve things, because lowering the highest fractional luminosity causes the lower limit of radius (mm) detectability to increase faster than the rate of decrease of the upper limit of radius (IR) detectability. This would produce too many belts at large radii, which are not observed. At the same time, we deem lowering the highest fractional luminosity to values below 10$^{-3}$ unrealistic, because 11/26 of our disks have fractional luminosities above this value.  

\textit{Log-uniform distribution of belt masses $M_{\rm belt}$}. We then assume a belt population that has a log-uniform distribution of belt masses $M_{\rm belt}$ rather than fractional luminosities. By belt mass we refer to the \textit{observable} dust mass of the belt as would be derived from millimeter observations \citep{Wyatt2008}, which differs from the true total belt mass which is dominated by unobservable bodies larger than mm/cm in size. The upper and lower boundaries of the distribution (0.5 and 10$^{-4}$ $M_{\oplus}$) are set to closely match the extremes in our observed sample.

For a log-uniform distribution of masses rather than fractional luminosities, given that in our model $f\propto M_{\rm belt}R^{-2}$, this means that belts at larger radii have lower fractional luminosities. In addition, for a log-uniform distribution of masses, the belt flux at a given wavelength $F_{\nu, \rm belt}$ depends on radius only through the belt temperature and not its mass (since $F_{\nu, \rm belt}\propto M_{\rm belt} B_{\nu}[T(R)]$ where $B_{\nu}[T(R)]$ is the Planck function). 

The result is that belts become undetectable only at large radii where the temperatures are too cold (Fig. \ref{fig:corner_uncorr_testass_1}, bottom left). The upper envelope is set by a line of constant flux equal to the detection threshold (dominated by 70$\mu$m observations), which follows $R\propto L_{\star}^{0.5}$ for a constant belt mass. Lowering the upper boundary of the distribution of belt masses has the effect of lowering this upper envelope of detectability; raising the lower boundary instead causes the selection probability to increase for each pixel, but without changing this upper boundary. 
The lower envelope of selected belts is determined solely by a belt's hard limit of resolvability with ALMA ($R>0\farcs028$).

This model assumption produces a large number of small belts, since these are much hotter and hence brighter than larger belts of the same mass. Such a large number of small belts is not observed, and this makes the simulated population very different from the observed one (Fig. \ref{fig:corner_uncorr_testass_1}, bottom right). This means that the chance of randomly drawing our observed power law parameters from an uncorrelated underlying population remains negligible.
Furthermore, an observed population of belts that is dominated by smaller belts would be inconsistent with the results of IR surveys \citep[e.g.][]{Sibthorpe2018}, which find a decrease in disk incidence at blackbody radii $<10$ au (at least for fractional luminosities above $10^{-5}$).

However, compared to the uniform fractional luminosity case, it is interesting to see that the upper envelope of detectability is now closer and has a slope more similar to the observed sample. We consider the observed lack of small disks and discuss its possible origin in \S\ref{sec:steadystateevol}.

\textit{Belt mass increasing with stellar luminosity, following the protoplanetary disk population}. Protoplanetary disk masses (or millimetre luminosities) are known to correlate with their host star's mass \citep{Andrews2013,Pascucci2016}. If this relation were to remain imprinted on planetesimal belts after protoplanetary disk dispersal, we would naively expect the same to apply here (neglecting any belt evolution during the main-sequence lifetime of the star, discussed in \S\ref{sec:steadystateevol}). We simulate a belt population where the belt masses are still created from a log-uniform distribution, but where now both the upper and lower boundaries of the distribution follow $M_{\rm belt}\propto M_{\star}^{\gamma}$, where $\gamma=1.7$ \citep[an average between the 1.5-1.9 range of values derived by][]{Pascucci2016}. The upper boundary of the mass distribution for the most luminous star and the lower boundary for the least luminous star are fixed to the extremes of our observed sample (0.5 and 10$^{-4}$ $M_{\oplus}$).

In Fig. \ref{fig:corner_uncorr_testass_2} (top) we find that this assumption causes little change compared to a mass distribution that is independent of stellar luminosity (Fig. \ref{fig:corner_uncorr_testass_1}, bottom). The dependence of the selection probability on stellar luminosity slightly steepens, as belts around more luminous stars will be intrinsically more massive. It also slightly increases the slope of the upper limit of detectability. This small change is overwhelmed by the still too large population of small belts, resulting in a simulated population that remains inconsistent with the observed data.

\textit{$R/R_{\rm BB}$ dependent on stellar luminosity}. The distribution of small dust grains in planetesimal belts, resolved by the \textit{Herschel} Space Observatory at 100 $\mu$m, show a trend where their ratio between resolved radii and blackbody radii ($R/R_{BB}$) decreases following a power-law as a function of stellar luminosity \citep{Pawellek2014}. Although the effect of observational bias on this result remains to be evaluated, we here assume the relation to be true and assess its impact on the detectability of a belt in $[R, L_{\star}]$ space. In particular, we take the best fit power law parameters $R/R_{\rm BB}=5.42L_{\star}^{-0.35}$ obtained for a 50\% astrosilicate + 50\% ice composition in the reanalysis of \citet{Pawellek2015}.

Going back to our original assumption of a belt population with log-uniform distribution of fractional luminosities, we find that introducing a $R/R_{\rm BB}$ dependence on $L_{\star}$ decreases the slope of the upper envelope of (70$\mu$m) detectability, making it nearly flat (Fig. \ref{fig:corner_uncorr_testass_2}, center left). This is because for the same radius, belts around less luminous stars are hotter and hence brighter. The lower envelope of (mm) detectability remains largely unchanged from the luminosity-independent $R/R_{\rm BB}$ case, due to a much weaker dependence of physical radius on temperature for a given flux detection threshold at mm compared to IR wavelengths. 

The predicted large scatter and, on average, larger radii than both the observed and other simulated populations mean that a $[R, L_{\star}]$ population with a luminosity-dependent $R/R_{\rm BB}$ as found by \textit{Herschel} studies is unable to explain the observed $R-L_{\star}$ relation. Furthermore, we do not find a significant $R/R_{\rm BB}$ relation to hold for our sample of belts resolved at mm wavelengths.

\textit{Fixed $R/R_{\rm BB}$, $\lambda_0$ and $\beta$.} Last, in an attempt to reduce the scatter in the simulated population, we fix $R/R_{\rm BB}$ and the modified blackbody parameters $\lambda_0$ and $\beta$ to a single value that is independent of $L_{\star}$, rather than varying them between the extremes observed in our population. We assume $R/R_{\rm BB}=2.87$ (the average value measured for our observed population), $\lambda_0=210$ $\mu$m \citep[the fiducial value of][]{Wyatt2008}, and $\beta=0.59$ \citep[the average best-fit value obtained from fitting the millimetre slope of millimetre-bright disks,][]{Macgregor2016a}. As shown in Fig. \ref{fig:corner_uncorr_testass_2} (bottom), we find no significant difference in the derived belt detectability in $[R, L_{\star}]$ space compared to the fiducial model assumptions, with the simulated intrinsic scatter remaining significantly larger than observed.

%% The following command ends your manuscript. LaTeX will ignore any text

\end{document}